\documentclass[aps,twocolumn,groupedaddress,nofootinbib]{revtex4-2}

\usepackage{formatAndDefs}
\usepackage[pdftex,pdfauthor={Julien Legendre},pdftitle={Operating conditions and thermodynamic bounds of dual radiative heat engines}]{hyperref}
\hypersetup{
    colorlinks,
    linkcolor={red!50!black},
    citecolor={blue!50!black},
    urlcolor={blue!80!black},
    filecolor=black
}

\begin{document}

\title{Operating conditions and thermodynamic bounds of dual radiative heat engines}

\author{Julien Legendre}
\affiliation{Univ Lyon, CNRS, INSA-Lyon, Université Claude Bernard Lyon 1, CETHIL UMR5008, F-69621, Villeurbanne, France}

\author{Pierre-Olivier Chapuis}
\affiliation{Univ Lyon, CNRS, INSA-Lyon, Université Claude Bernard Lyon 1, CETHIL UMR5008, F-69621, Villeurbanne, France}

\date{\today}

\begin{abstract}
    We propose a unified description of dual radiative heat engines (RHEs), consisting of two facing optoelectronic components (diodes) and capable of generating electrical power from heat. They can operate in three regimes depending on the applied biases, namely in thermoradiative-negative electroluminescent (TRNEL), thermoradiative-photovoltaic (TRPV) or thermophotonic regimes (TPX, consisting of a light-emitting diode and a PV cell). They have access to operating conditions that are unachievable by single RHEs such as thermophotovoltaic systems: at the radiative limit, TRNEL devices can reach the Carnot efficiency for any bandgap, while TPX devices achieve large power outputs by means of electroluminescent enhancement. The influence of non-radiative losses is also investigated, revealing the importance for dual engines to operate close to the radiative limit to clearly outperform single engines. Expressions of the maximum power output and related efficiency achieved by dual engines at the radiative limit are derived analytically, and reveal that the power output of TPX engines is not bounded. A comparison to usual efficiency bounds also highlights the impact of thermalisation losses, and the subsequent interest of spectral filtering to extend the operating region. This work provides common framework and guidelines for the study of RHEs, which represent a promising solution for reliable and scalable energy conversion.
\end{abstract}

\maketitle

\section{Introduction}

The conversion of heat into electrical power by means of solid-state heat engines \cite{Melnick2019}, such as thermoelectric \cite{Shi2020} and thermionic generators \cite{AbdulKhalid2016}, has gathered sizeable attention in recent decades due to their reliability and scalability. These engines also include optoelectronic systems, the most popular ones probably being photovoltaics for solar application, and thermophotovoltaics \cite{Burger2020,LaPotin2022,Tervo2022,Giteau2023}. In the latter case, the radiation comes from a hot emitter maintained at a high temperature by the heat supplied to the system. This gives access to a broad range of applications, for instance in latent heat thermophotovoltaic (TPV) batteries \cite{Datas2022}. But apart from photovoltaics and thermophotovoltaics, other optoelectronic systems are able to convert heat into electricity. One is the thermoradiative (TR) cell: as opposed to the photovoltaic (PV) cell, it is able to generate electrical power by emitting negative electroluminescent radiation to cold surroundings \cite{Strandberg2015,Pusch2019} (e.g. the night sky, outer space) or towards a cold absorber \cite{Santhanam2016}. TR cells, along with PV and TPV cells, are single radiative heat engines: they are heat engines in which one active component produces electrical power by either emitting or absorbing radiation \cite{Tervo2018}. Their typical electrical characteristic is provided in Appendix \ref{app:eleccharac}.

\begin{figure}
    \centering
    \begin{tikzpicture}[]
        \node (source)  [layer,color_heatsource,node distance=10mm,minimum height=0.2cm]               {\normalsize\textbf{Heat source} ($\Th$)};
        \node (machineh)[layer,color_hot,node distance=5mm,below=of source,minimum height=1cm]       {\normalsize\textbf{Hot optoelectronic component}\\($\muh,\Th$)};
        \node (machine) [layer,color_cold,node distance=7.5mm,below=of machineh,minimum height=1cm]    {\normalsize\textbf{Cold optoelectronic component}\\($\muc,\Tc$)};
        \node (sink)    [layer,color_heatsink,node distance=5mm,below=of machine,minimum height=0.2cm]{\normalsize\textbf{Heat sink} ($\Tc$)};
        
        \draw [ray,red!80] ($(machineh.south)+(-5mm,0)$) to node[auto,swap,color=black,xshift=-1mm]{\normalsize$(\hot{q},\Nh)$} ($(machine.north)+(-5mm,0)$);
        \draw [ray,blue!60!black] ($(machine.north)+(5mm,0)$) to node[auto,swap,color=black,xshift=1mm]{\normalsize$(\cold{q},\Nc)$} ($(machineh.south)+(5mm,0)$);

        \draw [arrow,red!80] ($(source.south)+(0,0)$)  to node[auto,xshift=1mm,color=black]{\normalsize$\mathsub{q}{source} $} ($(machineh.north)+(0,0)$);   
        \draw [arrow,blue!60!black] ($(machine.south)+(0,0)$)  to node[auto,xshift=1mm,color=black]{\normalsize$\mathsub{q}{sink}$} ($(sink.north)+(0,0)$);
        \draw [arrow,yellow!60!orange]  ($(machine.east)+(0mm,0mm)$) to node[xshift=9mm]{\large\color{black}$P$} ($(machine.east)+(14mm,0mm)$);
        \draw [arrow,yellow!60!orange]  ($(machine.east)+(6mm,0mm)$) |- ($(machineh.east)+(0mm,0mm)$);
        \draw [arrow,yellow!60!orange]  ($(machine.east)+(0mm,0mm)$) to ($(machine.east)+(7mm,0mm)$);
        \node at ($(machineh.east)+(4mm,4mm)$) {\large$\hot{P}$};
        \node at ($(machine.east)+(4mm,-4mm)$) {\large$\cold{P}$};
    \end{tikzpicture}
    \caption{Representation of the dual radiative engine (here in a thermophotonic configuration). It consists of two optoelectronic components which exchange electroluminescent radiation. By thermally connecting them to a heat source and a heat sink, the heat supplied can be converted into electrical power.}
    \label{fig:dualengine}
\end{figure}
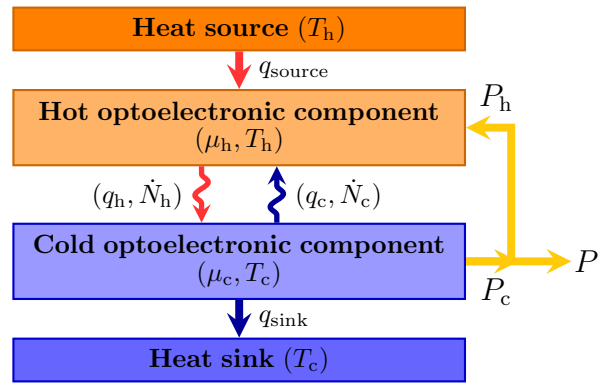

It is also possible to couple two different optoelectronic components into a dual radiative heat engine (see Fig. \ref{fig:dualengine}). One such dual engine is the thermoradiative-photovoltaic (TRPV) device \cite{Liao2019,Tervo2020}, in which a hot TR cell is coupled to a cold PV cell: both components are then able to generate electrical power, although production by the TR cell limits its emission due to negative electroluminescence, and therefore reduces the PV cell production. Nonetheless, having two components provides better control of the operating conditions. Thermophotonics \cite{Harder2003,McSherry2019,Sadi2022} has also gathered interest recently. In a thermophotonic (TPX) device, the hot emitter is a light-emitting diode (LED). While an LED consumes electrical power, it can emit electroluminescent radiation towards the PV cell with an above-unity wall-plug efficiency (i.e. emitting more power than it consumes), enabling a significant increase in power output \cite{Santhanam2012}. Recently, TPX devices have mostly been studied in near-field operation, as near-field radiation further increases the power output \cite{Zhao2017,Legendre2022,Legendre2025} and limits the impact of non-radiative losses \cite{Legendre2022a}.

Although the aforementioned systems are all radiative heat engines and share the same core physical laws, they are studied independently of one another in the literature. In this manuscript, we provide a unified overview of the performance of dual radiative engines, for which every device previously mentioned corresponds to a specific operating regime. This analysis is performed first at the radiative limit, then with non-radiative losses included, highlighting the respective merits of each engine and providing a fair comparison of their performance. It also clarifies their capabilities in comparison to single radiative engines. We particularly focus on the maximum achievable power and efficiency, for which we are able to derive closed-form expressions. The comparison to usual limits reveals the significant influence of thermalisation, making spectral filtering an attractive solution to achieve better efficiency.

\section{Dual radiative heat engines}\label{sec:etaP}

In the following, we assume that the dual engine operates in the far field, with the optoelectronic component on the cold side being kept at ambient temperature ($\Tc=300$ K). To maximise the achievable power output, the emission of radiation is assumed to follow the generalised Planck law, which corresponds to the extension of the usual Planck law for non-thermal (here, electroluminescent) radiation \cite{Wurfel1982,DeVos1992}. The bandgaps of the two components are assumed to be equal. For TPX systems, this condition has been shown to maximise power \cite{Zhao2018}, as long as the influence of Urbach tails is neglected (see \cite{Legendre2023}, pp. 125-130). Furthermore, the radiation exchanged below the bandgap is neglected to obtain an upper bound of the efficiency. Thus, the emitted photon flux density $\dot{N}$ and the related heat flux density $q$, respectively expressed in photons and energy per unit time and area, can be written as
\begin{subequations}
    \begin{align}
        \dot{N}_i&=\frac{1}{4\pi^2c^2\hbar^3}\int_{\Eg}^{\infty} \frac{E^2}{\exp\left(\frac{E-\mu_i}{\kb T_i}\right)-1}dE,\\
        q_i&=\frac{1}{4\pi^2c^2\hbar^3}\int_{\Eg}^{\infty} \frac{E^3}{\exp\left(\frac{E-\mu_i}{\kb T_i}\right)-1}dE,
    \end{align}
\end{subequations}
where $i$ relates to the emitting body (either “h” or “c”) with bandgap energy $\Eg$ and temperature $T_i$. $\mu_i$ is the chemical potential of the emitted radiation, which must remain strictly smaller than $\Eg$ and is assumed to be related to the voltage $U_i$ applied to the component through $\mu_i=eU_i$ \cite{Wurfel1982,Callahan2021,Legendre2025}, $e$ being the elementary charge. The above integrals can be expressed analytically using polylogarithms \cite{Green2012a}, leading to
\begin{subequations}\label{eq:polylog}
    \begin{align}
        \begin{split}
            \dot{N}_i&=\frac{(\kb T_i)^3}{4\pi^2c^2\hbar^3}\bigl(e_{g,i}^2\polylog_1(e^{-x_{i}})+2e_{g,i}\polylog_2(e^{-x_{i}})\\
            &\hphantom{=\frac{(\kb T_i)^3}{4\pi^2c^2\hbar^3}}+2\polylog_3(e^{-x_{i}})\bigr),
        \end{split}
        \\
        \begin{split}
            q_i&=\frac{(\kb T_i)^4}{4\pi^2c^2\hbar^3}\bigl(e_{g,i}^3\polylog_1(e^{-x_{i}})+3e_{g,i}^2\polylog_2(e^{-x_{i}})\\
            &\hphantom{=\frac{(\kb T_i)^4}{4\pi^2c^2\hbar^3}}+6e_{g,i}\polylog_3(e^{-x_{i}})+6\polylog_4(e^{-x_{i}})\bigr), 
        \end{split}
    \end{align}
\end{subequations}
where $\polylog_n$ is the $n$th-order polylogarithm, $e_{g,i}=\Eg/\kb T_i$ and $x_{i}=(\Eg-\mu_i)/\kb T_i$. Finally, to obtain an upper bound of the dual engine performance, we assume that any active component operates at the radiative limit (i.e. with a perfect conversion of photons into charges and vice-versa, or said otherwise without any non-radiative losses). The electrical power generated by a component $i$ facing a component $j$ is then
\begin{equation}\label{eq:power_i}
    P_i=U_i\cdot e(\dot{N}_j-\dot{N}_i)=\mu_i(\dot{N}_j-\dot{N}_i).
\end{equation}
The total power output is therefore
\begin{equation}\label{eq:power_tot}
    P=(\muc-\muh)(\Nh-\Nc).
\end{equation}
Since $\mathsub{q}{source}=\hot{q}-\cold{q}+\hot{P}$, the overall heat engine efficiency is
\begin{equation}\label{eq:efficiency}
    \eta=\frac{P}{\mathsub{q}{source}}=\frac{(\muc-\muh)(\Nh-\Nc)}{\hot{q}-\cold{q}-\muh(\Nh-\Nc)}.
\end{equation}

\begin{figure}
    \centering
    \includegraphics{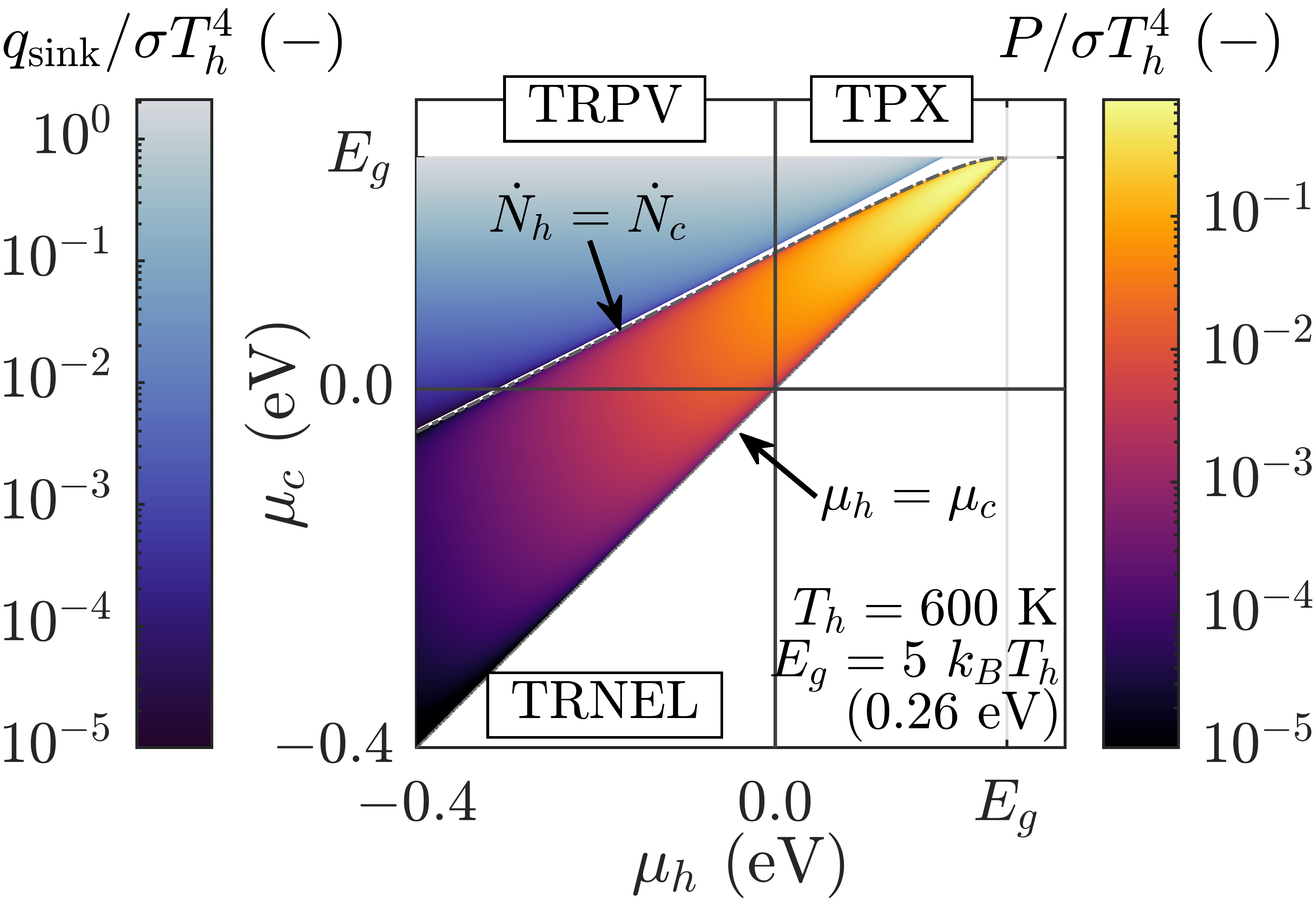}
    \caption{Performance of dual radiative systems at the radiative limit and for $\Eg=5\kb \Th=0.26$ eV. This performance is quantified by the electrical power output $P$ for heat engines (central area, right colour scale), and by the heat extracted from the heat sink $\mathsub{q}{sink}=\hot{q}-\cold{q}-\cold{P}$ for heat pumps used for cooling purposes (upper-left area, left colour scale).}
    \label{fig:mumu}
\end{figure}

We start by studying how the power output of the dual engine varies in the ($\muh$,$\muc$) plane. The results obtained considering $\Th=600$ K and $\Eg=5\kb \Th$ are illustrated in Fig. \ref{fig:mumu}. Although we are mostly interested in heat engine operation, the dual system can also operate as a heat pump \cite{Santhanam2012,Radevici2019}, and we indicate the cooling power on a blue colour scale. Note that such high cooling powers can only be achieved because below-bandgap radiation is neglected \cite{Chatelet2024}.

For the bandgap selected, the dual system is able to operate both as a heat engine and a heat pump in three of the four quadrants: namely, the TPX quadrant ($\muh>0,\muc>0$), the TRPV quadrant ($\muh<0,\muc>0$) but also the “TRNEL” quadrant ($\muh<0,\muc<0$), in which a cold negative electroluminescent (NEL) diode consumes electrical power to limit the radiation sent to the hot facing TR cell. The basic principle of TRNEL operation is similar to that of TPX operation: by supplying one component with electricity, the power generated by the other component get enhanced. The only difference is that TRNEL devices rely on negative (rather than positive) electroluminescence. For more details on TRNEL operation, the corresponding energy diagram can be found in Appendix \ref{app:eleccharac}. To the best of our knowledge, TRNEL devices have never been mentioned in the literature. Going back to Fig. \ref{fig:mumu}, both the power output and the cooling power appear to increase with $\muc$, while $P$ also increases with $\muh$: the maximum power point (MPP) is therefore located in the TPX quadrant, while the maximum cooling power is reached in TRPV operation for $\muc\rightarrow\Eg$, where it converges towards a constant value \cite{Zhao2020}. Additionally, note that for low $\Eg$, the TPX device becomes unable to perform cooling (see Appendix \ref{app:lowEg}.).

\begin{figure}
    \centering
    \includegraphics{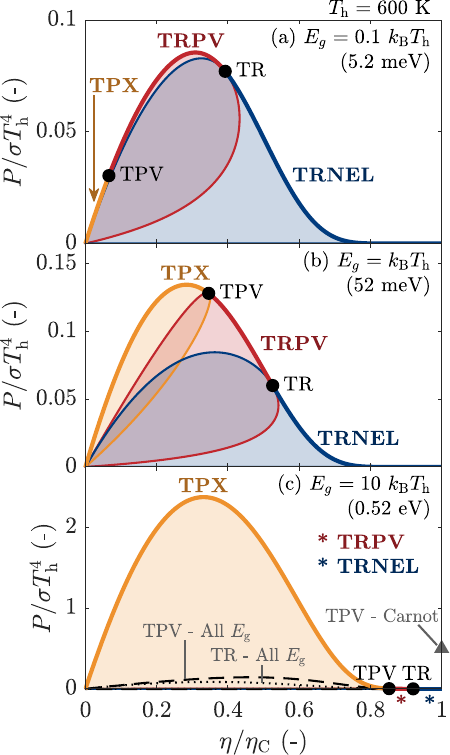}
    \caption{$\eta-P$ plots of dual radiative engines at the radiative limit, for $\Th=600$ K and for various bandgaps. The coloured areas represent the $\eta-P$ couples achievable by each dual radiative engine, while the full lines correspond to the envelope of these areas. Power output is normalised by the blackbody emissive power at $\Th$ to make the comparison between systems with different heat source temperatures easier. Efficiency is given relative to the Carnot efficiency: it therefore corresponds to the exergy efficiency or second-law efficiency, and is bounded by one.}
    \label{fig:etaP600}
\end{figure}

Note how the dual system does not switch directly from heat engine to heat pump operation as $\mu$ varies. Indeed, there is a narrow region in-between where the system is able neither to generate electrical power nor to cool the cold source. This gap can especially be observed when $\Eg-\muh$ becomes lower than $\kb \Th$, but globally narrows down as the bandgap increases: in the limit of infinite bandgaps, the two operating regions become adjacent as $\hot{q}=\cold{q}$ and $\Nh=\Nc$ are equivalent. In this situation, the term related to the lowest-order polylogarithm dominates in the expressions provided in Eq. \eqref{eq:polylog}. By linearising $\polylog_1$, we obtain that the transition from one region to the other, which corresponds to open-circuit conditions since currents are then equal to zero, occurs approximately for
\begin{equation}\label{eq:oc}
    \muc=\frac{\Tc}{\Th}\muh+\left(1-\frac{\Tc}{\Th}\right)\Eg+\kb \Tc\ln\left(\frac{\Th}{\Tc}\right),
\end{equation}
this linearised expression being a very good approximation as long as $\Eg-\mu_i\gg \kb T_i$. The other side of the power production region is simply delimited by the condition $\muh=\muc$. These two expressions have already been mentioned for TPX devices in \cite{Legendre2022}.

To quantify and compare the performance of TPX, TRPV and TRNEL devices, we provide in Fig. \ref{fig:etaP600} the $\eta-P$ plots obtained by varying both $\muh$ and $\muc$, considering three different bandgaps and a heat source temperature of 600 K. For each device, the shaded area corresponds to achievable operation, while the full line is the envelope of this area. It is first interesting to notice that while the shape of the admissible $\eta-P$ area changes significantly when considering each engine individually, it remains mostly unchanged with varying bandgaps for the full dual radiative engine. For any of the bandgaps considered, the efficiency at maximum power remains mostly constant (between 28\% and 34\% of the Carnot efficiency $\etaC=1-\Tc/\Th$), while the maximum efficiency is always $\etaC$ and is reached at zero power.

If we now compare the different regimes, the TRNEL appears to be the one achieving the highest efficiencies. This is enabled by the strong decay of the entropy production when $\mu_{h/c}\rightarrow-\infty$, which decreases much faster than the power output. In open-circuit conditions, which are reached roughly for $\muh/\kb\Th=\muc/\kb\Tc$ when $\mu_{h/c}\rightarrow-\infty$, this leads the efficiency to reach the Carnot limit. This can be proven using that in such conditions, the ratio $(\hot{q}-\cold{q})/(\Nh-\Nc)$ is independent of the chemical potentials and can thus be neglected in comparison to $\muh$ in Eq. \eqref{eq:efficiency}. However, the advantage of TRNEL operation shrinks for large bandgap: as $\Eg\rightarrow\infty$, all radiative engines are able to approach the Carnot efficiency in open-circuit conditions. This has already been demonstrated for TR \cite{Strandberg2015,Pusch2019} and TPV \cite{Roux2024}, but can in fact be shown for any radiative engine. In the limit of an infinite bandgap $(\hot{q}-\cold{q})/(\Nh-\Nc)\rightarrow \Eg$, leading $\eta$ to be expressed as $(\muc-\muh)/(\Eg-\muh)$. Using Eq. \eqref{eq:oc}, and keeping in mind that $\Eg-\muh\gg \kb \Th$, we obtain that the efficiency in open-circuit conditions equals $\etaC$.

To maximise the power output, TPX is almost always the best candidate, TRPV becoming optimal only for very low bandgap energies (here, for $\Eg<\kb \Th\approx 0.05$ eV) which are difficult to achieve in practice. This remains true for lower or higher heat source temperatures (see Appendix \ref{app:impactT}). TPX devices generally outperform other radiative engines in terms of power output because the hot emitter operates as an LED: it is therefore able to largely enhance its emission by electroluminescence, increasing consequently the various energy flows in the system. In fact, the power output of TPX devices is unbounded at the radiative limit, the maximum power diverging when $\Eg\rightarrow\infty$ (see Appendix \ref{app:impactEg}). For high enough bandgaps, $\mathsub{P}{max}$ can thus exceed $\sigma \Th^4$ (see Fig. \ref{fig:etaP600}c). This is impossible for single engines, TPV power output being for instance bounded by $P=\etaC\sigma\Th^4$ , and even lower in practice: the best performances achieved for any bandgap, represented by a dashed line in Fig. \ref{fig:etaP600}c, are indeed far from the $P/\sigma\Th^4=\etaC=0.5$ bound. In addition, since dual engines perform well even for large $\Eg$, they are not restricted to mid-infrared bandgaps for low $\Delta T$, and can also be made of near-infrared or visible diodes which typically operate closer to the radiative limit. This, along with the significant power enhancement previously mentioned, represent some of the main interests of dual radiative engines.

While TRNEL devices maximise efficiency and TPX devices power, TRPV systems can in some conditions provide interesting trade-offs between power and efficiency (see Fig. \ref{fig:etaP600}b). However, this can only be observed for low bandgaps (a few $\kb \Th$ at most) which are hardly achievable for the heat source temperature considered.

\begin{figure}
    \centering
    \includegraphics{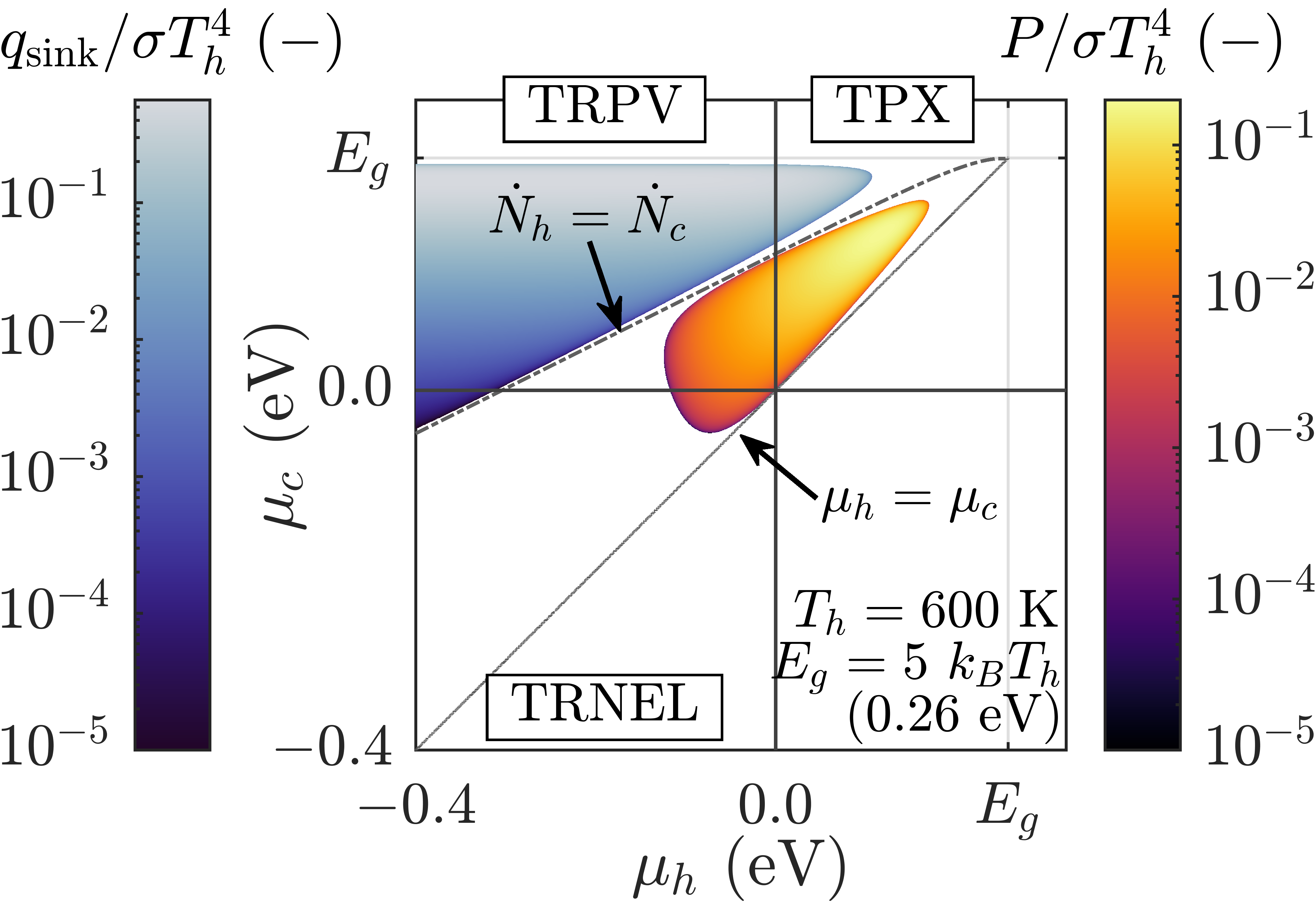}
    \caption{Performance of dual radiative systems operating as heat engines or heat pumps, for $\mathrm{QE}=0.9$ and $\Eg=5\kb \Th=0.26$ eV.}
    \label{fig:mumu_QE}
\end{figure}

\begin{figure}
    \centering
    \includegraphics{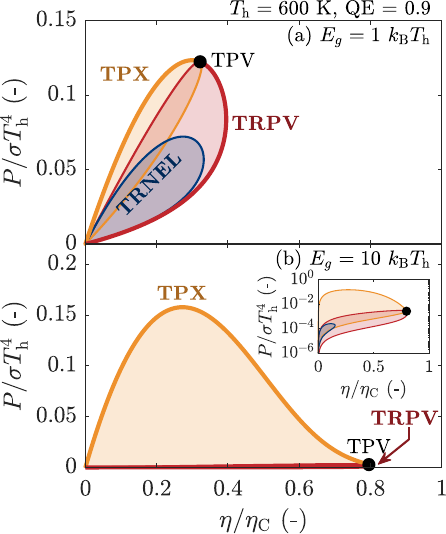}
    \caption{$\eta-P$ plots obtained for dual radiative engines, for $\Th=600$ K and for various bandgaps. A quantum efficiency of 0.9 is considered.}
    \label{fig:qe}
\end{figure}

\begin{figure}
    \centering
    \includegraphics{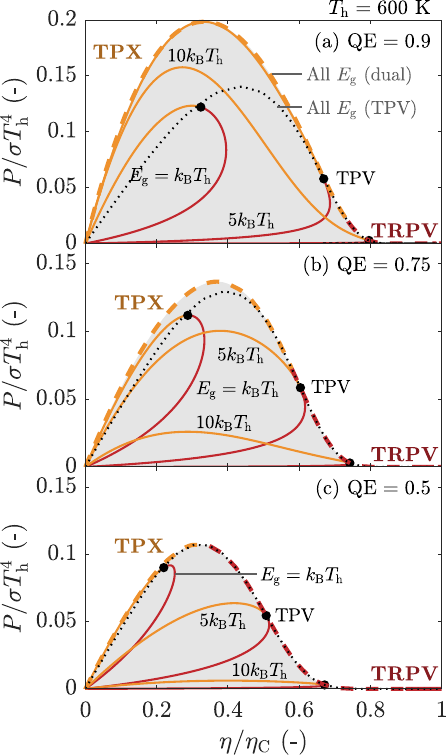}
    \caption{Dual radiative engine $\eta-P$ envelopes obtained for varying bandgap energy $\Eg$, depending on the quantum efficiency considered. The thin full lines in colour are the envelopes obtained for dual engines at a given $\Eg$. The thick coloured dashed line and the black dotted line are the envelopes obtained by considering the whole $\Eg$ space for dual engines and TPV engines, respectively.}
    \label{fig:envelopeQE}
\end{figure}

\section{Impact of non-radiative losses on $\eta-P$ characteristics}\label{sec:nonrad}

The results above are obtained at the radiative limit, and therefore provide an upper bound for dual radiative engines' performance. Obviously, if non-radiative losses are included, performance will be worsened: we aim to give here some insights about the impact of such losses. To do so, we manually set the quantum efficiency (QE), defined as the fraction of recombinations being radiative \cite{Harder2003}, to 0.9 for both components of the dual engine. This means that 10\% of the total recombination rate can be attributed to non-radiative losses. Since the non-radiative generation rate must balance the non-radiative recombination rate at equilibrium, we obtain a new expression of the power generated by an optoelectronic component \cite{Legendre2022}:
\begin{equation}\label{eq:qe}
    P_i=\mu_i\left(\dot{N}_j-\dot{N}_i-\frac{1-\mathrm{QE}}{\mathrm{QE}}\left(\dot{N}_i-\dot{N}_i(\mu_i=0)\right)\right).
\end{equation}
Note that the definition of the QE used here corresponds to the one usually considered in the LED community, and should not be confused with its PV counterpart which corresponds to the conversion efficiency of photons into charges. For a 600-K heat source temperature, the addition of non-radiative events strongly modifies the performance of the engine, as shown in Fig. \ref{fig:mumu_QE} where the variations of the electrical power output and the cooling power with the chemical potentials are depicted. The resulting $\eta-P$ plots are drawn in Fig. \ref{fig:qe}, and reveal several major differences in comparison to the results obtained at the radiative limit. First, TRNEL devices have significantly lost interest, no longer being able to reach Carnot efficiency. Worse, their envelope is now within that of TRPV devices for both bandgaps considered. Therefore, TRNEL systems are of interest only when operating extremely close to the radiative limit: for TRNEL maximum efficiency to exceed that of TRPV, QE should be greater than 0.98 for $\Eg=\kb\Th$, and even greater than 0.9999999 for $\Eg=10\kb\Th$. For more feasible values of QE, the maximum efficiency of the dual radiative engine is reached in TRPV regime. Second, the maximum efficiency is now highly dependent on the bandgap: being only 40\% of Carnot efficiency for $\Eg=\kb\Th$, it goes up to 80\% of $\etaC$ for $\Eg=10\kb\Th$. Finally, the power output variation with the bandgap is much weaker. In fact, as long as $\mathrm{QE}<1$, there is an optimal bandgap that allows the power output to be maximised: the optimum bandgap is close to $\Eg=5\kb\Th$ for $\Th=600$ K and $\text{QE}=0.9$, and decreases for the two lower QEs considered. The complete variations of the optimum bandgap with QE are further analysed in Appendix \ref{app:optEg}. 

Because the maximum power no longer diverges with $\Eg$, we can trace the $\eta-P$ envelope of dual engines obtained over the whole bandgap space, indicating the best operating conditions achievable for a certain quantum efficiency. These are illustrated in Fig. \ref{fig:envelopeQE} for three different quantum efficiencies, along with the TPV envelope in dashed line for comparison. In all three scenarios, the performance of dual radiative engines at high efficiency is similar to those of TPV engines, with maximum efficiency approaching the Carnot limit for infinitely large bandgap. This originates from the fact that $\Nh/\Nc\rightarrow\infty$ when $\Eg\rightarrow\infty$, making the influence of QE in Eq. \eqref{eq:qe} negligible for very high bandgaps. When operating in TPX regime, dual engines can reach higher power outputs than TPV even when non-radiative losses are included. However, quantum efficiencies of at least 0.75 are required to obtain a noticeable improvement compared to TPV. In addition, the optimal bandgap (in terms of power output) remains quite low even for $\mathrm{QE}=0.9$, being equal to $5\kb\Th=0.26$ eV. To make materials with bandgaps in the near-infrared and the visible range attractive, larger QEs should be achieved. For instance, for $\Eg=20\kb\Th\approx 1$ eV, a QE of at least 0.94 is required to deliver an electrical power output larger than that of TPV. However, the dual engine maximum power quickly increases for larger QEs, rising to $6\sigma\Th^4$ at the radiative limit for this bandgap energy.

As a side note, including non-radiative losses causes a mismatch between the currents $\hot{J}$ and $\cold{J}$ of the two optoelectronic components. Since voltages are mismatched too, the two components cannot be directly bound electrically if both have the same area, and additional electronics is necessary to make the engine work. Otherwise, it is possible to design engines with components with mismatched areas \cite{Zhao2019} or bandgaps \cite{Yang2024a} to make them self-sustaining.

\section{Performance bounds at maximum power}

We now return to the analysis of the system at the radiative limit, to understand what are the performance bounds of dual radiative engines. The study of $\eta-P$ plots in Fig. \ref{fig:etaP600} directly highlights that the maximum efficiency of dual radiative engines is always the Carnot efficiency, reached for zero power output. Regarding the MPP, it was observed that $\mathsub{P}{max}$ increases significantly with $\Eg$ while $\mathsub{\eta}{MPP}$ varies only slowly, but no analytical expressions of these quantities have yet been formulated. Therefore, in the following, we focus on the analytical derivation of $\mathsub{P}{max}$ and $\mathsub{\eta}{MPP}$. To do so, we consider that $\Eg\gg\kb\Th$, since this allows reaching the largest possible power output (see Appendix \ref{app:impactEg}). By doing so, the $\polylog_1$ term dominates the expressions provided in Eq. \eqref{eq:polylog}. The following derivations, which are summarised in Appendix \ref{app:analyticBroad}, indicate that maximum power output is reached for $\mu_i=\Eg-\ln(2)\kb T_i$, where it is expressed as
\begin{equation}\label{eq:Pmax_infty}
    \mathsub{P}{max}=\frac{1}{\hbar}\left(\frac{\ln(2)\Eg\kb(\Th-\Tc)}{2\pi c\hbar}\right)^2.
\end{equation}
It varies quadratically with the bandgap energy: unlike PV or TPV systems, the power output of dual radiative engines is not bounded at the radiative limit, as mentioned in Section \ref{sec:etaP}. $\mathsub{P}{max}$ also depends quadratically on the temperature difference, similarly to thermoelectric engines \cite{Apertet2012}. These dependences of $\mathsub{P}{max}$ on $\Eg$ and $\Delta T$ were already pointed out in \cite{Zhao2020}, although without a complete closed-form expression. Since $\Eg\gg \kb T_i$, both chemical potentials are positive and the maximum power is reached in the TPX regime, consistently with the results from Fig. \ref{fig:etaP600}.

\begin{figure}
    \centering
    \includegraphics{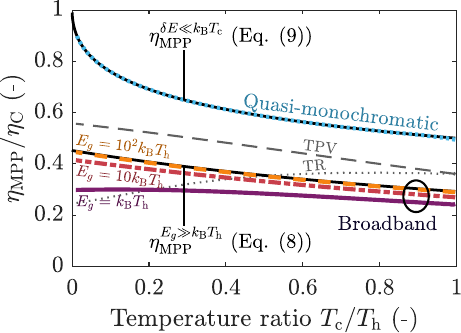}
    \caption{Variation of the efficiency at maximum power $\mathsub{\eta}{MPP}$ with the operating temperature ratio $\Tc/\Th$. For simplicity, only the TPX quadrant has been considered. The black lines represent the analytical expressions, while the coloured lines correspond to numerical results, either in the quasi-monochromatic case (in blue) or in the broadband case for various bandgaps (from purple to yellow). For comparison, the efficiencies at maximum power achieved by single radiative engines at the optimum bandgap and for broadband radiation are drawn in grey.}
    \label{fig:etaT}
\end{figure}

The efficiency at maximum power can be written as
\begin{equation}\label{eq:etaMPP_infty}
    \eta_{\mathrm{MPP}}^{\Eg\gg\kb\Th}=\frac{\etaC}{1+(2-\etaC)\chi},
\end{equation}
$\chi$ being a constant equal to $\frac{1}{2}(\frac{1}{6}(\frac{\pi}{\ln(2)})^2-1)\approx 1.21$. The temperature variation of $\eta_{\mathrm{MPP}}^{\Eg\gg\kb\Th}$ is provided in Fig. \ref{fig:etaT} (black line), and matches well the numerical results obtained for bandgaps larger than $100\kb\Th$. It also gives a good estimate of the efficiency obtained for standard bandgaps, as long as $\Eg\gg \kb\Th$: for $\Th=600$ K, $\eta_{\mathrm{MPP}}^{\Eg\gg\kb\Th}=17.7$\% while $\eta_{\mathrm{MPP}}^{\mathrm{1\; eV}}=$ 17.1\% and $\eta_{\mathrm{MPP}}^{\mathrm{0.52\; eV}}=16.6$\%. In addition, because maximum power diverges when $\Eg\rightarrow\infty$, below-bandgap radiation would be negligible in comparison to $P$, being independent of the chemical potential. Consequently, Eq. \eqref{eq:etaMPP_infty} holds even when below-bandgap radiation is included, and corresponds to the analytical extension of the Shockley-Queisser limit \cite{Shockley1961} for dual radiative engines.

\begin{figure}[b]
    \centering
    \includegraphics{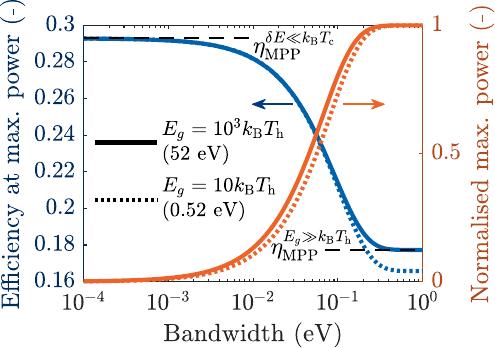}
    \caption{Variation of the maximum power (in blue) and related efficiency (in orange) for varying spectral bandwidth, considering $\Th=600$ K. For simplicity, only the TPX quadrant has been considered.}
    \label{fig:bandwidth}
\end{figure}

To better understand how efficient dual radiative engines are at maximum power, one can compare Eq. \eqref{eq:etaMPP_infty} with classical upper bounds for $\mathsub{\eta}{MPP}$. The Novikov-Curzon-Ahlborn efficiency $\mathsub{\eta}{NCA}=1-\sqrt{\Tc/\Th}$ \cite{Novikov1957,Curzon1975} and the Schmiedl-Seifert efficiency $\mathsub{\eta}{SS}=2\etaC/(4-\etaC)$ \cite{Schmiedl2008} are respectively efficiency bounds for endoreversible and exoreversible engines such as thermoelectric generators \cite{Apertet2012a}. For the temperatures previously considered, both efficiencies are close to 29\%, hence 11 percentage points higher than $\eta_{\mathrm{MPP}}^{\Eg\gg\kb\Th}$. This significant difference, which highlights the presence of additional losses in radiative engines, can be attributed to thermalisation losses - that is, to the fraction of radiative energy exchanged which is useless to optoelectronic conversion. To verify this, we now consider the radiation to be quasi-monochromatic around $\Eg$ (i.e. $\Eg\le \hbar\omega\le \Eg+\delta E$ with $\delta E\ll \kb\Tc$) so that thermalisation losses become negligible. As demonstrated in Appendix \ref{app:analyticNarrow}, this leads to
\begin{equation}
    \eta_{\mathrm{MPP}}^{\delta E\ll\kb\Tc}=1-\sqrt{\frac{\Tc}{\Th}},
\end{equation}
exactly the Novikov-Curzon-Ahlborn efficiency. $\delta E$ being the radiation spectral bandwidth, the maximum power output is then
\begin{equation}
    \mathsub{P}{max}=\frac{1}{\hbar}\left(\frac{\Eg\sqrt{\kb}(\sqrt{\Th}-\sqrt{\Tc})}{2\pi c\hbar}\right)^2\delta E,
\end{equation}
an expression similar to Eq. \eqref{eq:Pmax_infty}. Since $P$ goes to zero as $\delta E\rightarrow 0$, there is a trade-off between power and efficiency as a function of bandwidth, as illustrated in Fig. \ref{fig:bandwidth}: to achieve non-zero output power, the efficiency must fall below the usual Novikov-Curzon-Ahlborn bound. It is noteworthy that the efficiency starts to decrease for bandwidths as low as few meV (corresponding to a quality factor $Q=\Eg/\delta E$ close to 100 for $\Eg=0.52$ eV), while reaching the broadband limit for a bandwidth of few tenths of eV (i.e. for $Q\approx 1$ considering $\Eg=0.52$ eV). If the efficiency at maximum power is too low for a given application, two main leverages are thus available to increase it, although at the expense of power: decrease the radiation bandwidth, or change $\muh$ and $\muc$ to move in the broadband $\eta-P$ plots \cite{Giteau2023}. The interest of the latter solution can be inferred from the various $\eta-P$ plots provided above (see Figs. \ref{fig:etaP600}, \ref{fig:qe}, \ref{fig:envelopeQE}). The performances resulting from narrowing the radiation bandwidth by means of spectral filtering (see Fig. \ref{fig:filtering}) are examined below.

\section{Modification of the achievable operating conditions under spectral filtering}

\begin{figure}
    \centering
    \begin{tikzpicture}[]
        \node (machineh)[layer,color_hot,node distance=5mm,minimum height=1cm]       {\normalsize\textbf{Hot optoelectronic component}\\($\muh,\Th$)};
        \node (machine) [layer,color_cold,node distance=15mm,below=of machineh,minimum height=1cm]    {\normalsize\textbf{Cold optoelectronic component}\\($\muc,\Tc$)};

        \draw [dashed] ($0.5*(machineh.south west)+0.5*(machine.north west)$) -- ($0.5*(machineh.south east)+0.5*(machine.north east)$) node [anchor = north east] {filter};
        
        \draw [arrow,black] ($(machineh.south)+(-5mm,0)$) to ($0.5*(machineh.south)+0.5*(machine.north)+(-1mm,0)$) to node[anchor = east,color=black]{\normalsize$q_{\Eg\le E\le \Eg + \delta E}$} ($(machine.north)+(3mm,0)$);
        \draw [arrow,black] ($(machineh.south)+(-3mm,0)$) to ($0.5*(machineh.south)+0.5*(machine.north)+(1mm,0)$) to node[anchor = west,color=black, xshift = 1mm]{\normalsize$q_{E>\Eg + \delta E}$} ($(machineh.south)+(5mm,0)$);
    \end{tikzpicture}
    \caption{Representation of spectral filtering in dual radiative engines.}
    \label{fig:filtering}

    \includegraphics{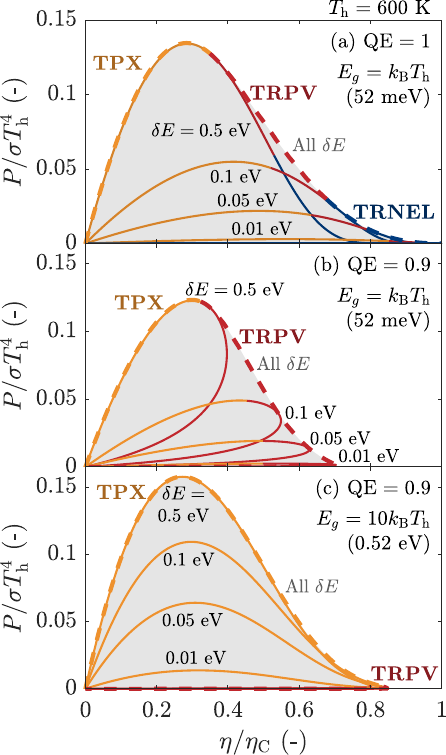}
    \caption{Dual radiative engine $\eta-P$ envelopes obtained for varying spectral bandwidth $\delta E$. The thin full lines are the envelopes obtained for dual engines at a given $\delta E$, while the thick dashed line corresponds to the envelope obtained by considering the whole $\delta E$ space.}
    \label{fig:envelopeBandwidth}
\end{figure}

We show in Fig. \ref{fig:envelopeBandwidth} how the $\eta-P$ envelope of dual engines varies with $\delta E$. In addition, the complete set of operating conditions achievable by varying $\delta E$ is represented by the grey area. Three different scenarios are considered to highlight the variability of the influence of spectral filtering on the achievable operating conditions.

The envelopes obtained at the radiative limit for a bandgap of $\kb\Th$ are shown in panel (a), the impact of spectral filtering being mostly similar for other bandgaps. In this scenario, filtering gives access to new operating conditions in the high-efficiency region. For $P/\sigma \Th^4<0.03$, this allows increasing the efficiency by up to 10 percentage points. Conversely, considering $\eta/\etaC=0.7$, normalised power output rises from 0.005 to 0.025: in other words, spectral filtering can limit the power loss undergone in high-efficiency operation. The effect is even stronger when decreasing the quantum efficiency (QE) to 0.9, as depicted in panel (b): an efficiency increase of the order of 25 percentage points can be reached for $P/\sigma \Th^4=0.03$, and goes up to 45 percentage points for $P/\sigma \Th^4=0.01$. In this case, filtering gives access to efficiencies that are unachievable with broadband radiation, and increases the maximum efficiency by 30\% compared to the broadband case. However, filtering only brings significant benefit when $\Eg$ remains smaller than a few $\kb\Th$, or at the radiative limit. Otherwise, reducing the bandwidth mostly makes the envelope shrink, the only benefit being a minor increase of the efficiency achieved for powers close to zero. This is clearly illustrated in panel (c), obtained for $\text{QE}=0.9$ and $\Eg=10\kb\Th$.

\section{Conclusion}

In conclusion, we have studied the power output and efficiency achievable by dual radiative heat engines. This unified description illuminates the similarities and respective merits of each operating regime. TRNEL devices are found to reach Carnot efficiency for any bandgap at the radiative limit, but see their performance decay quickly as non-radiative losses increase. Then, maximum efficiency is rather achieved in TRPV regime. For maximising power output, the TPX regime is almost always the best and offers the broadest range of operating conditions for bandgaps over a few $\kb\Th$, both at the radiative limit and when non-radiative losses are included. In comparison to TPV or TR systems, the use of dual engines significantly improves the maximum efficiency of low-bandgap devices and the maximum power of higher-bandgap devices.

We have derived analytical expressions for the maximum power and related efficiency in the $\Eg\rightarrow\infty$ limit, where power is maximised. This has revealed that the maximum power of dual engines is actually not bounded, and that the efficiency is several percentage points below usual efficiency limits due to thermalisation losses. We have also highlighted that spectral filtering can mitigate part of the power loss when high efficiencies are targeted. 

In this work we have determined the performance of dual radiative engines in a best-case scenario, and highlighted how such performance changes in the presence of non-radiative losses. In order to model more practical devices, below-bandgap radiation, more advanced models of non-radiative losses and resistive losses shall be included, as well as thermal resistance effects at the thermostats which reduce the operating temperature difference $\Th-\Tc$ (see \cite{Legendre2023}, pp. 76-85). An interesting work could be to study the potential of multijunctions for dual radiative engines. In addition, it would be valuable for the development of dual engines to quantify the impact of high-temperature operation on the QE as well as to develop solutions to mitigate it. Although we have briefly studied this recently \cite{Legendre2025}, it is still a fairly unexplored area. Finally, it would be worth investigating how the performance of such engines changes in the near field, where radiative emission exceeds the generalised Planck law \cite{Zhao2017,Legendre2022,Legendre2023}.

\begin{acknowledgments}
    This work has received funding from the European Union's Horizon 2020 research and innovation programme under Grant Agreement No. 951976 (TPX-Power project). The authors thank T. Châtelet, P. Kivisaari, O. Merchiers, J. Oksanen and J. van Gastel.
\end{acknowledgments}

\appendix

\section{Electrical characteristic and energy diagram of individual optoelectronic components}\label{app:eleccharac}

\begin{figure}[!t]
    \centering
    \begin{tikzpicture}
        \begin{axis}[
            ticks=none,
            xlabel={$U$},
            ylabel={$J$},
            axis lines=middle,
            every axis x label/.style={at={(current axis.right of origin)},anchor=west},
            every axis y label/.style={at={(current axis.above origin)},anchor=south},
            height=5.8cm,
            width=7.2cm,
            domain=-4.5:4.5,
            samples=200,
            xmin=-4.5, xmax=4.5,
            ymin=-2.5, ymax=2.5
        ]
  
        \addplot+[no markers,very thick,red] {-0.1+(-1.3+2)*exp(\x/1.5)};
        \addplot+[no markers,very thick,blue] {-2+(-1.75+2)*exp(\x)};

        \draw[fill = blue, opacity = 0.2] (0,0) rectangle (axis cs: 1.4,-0.986);
        \node at (axis cs: 1.4/2,-0.986/2) [anchor = center] {$\mathsub{P}{PV}$};
        
        \node[circle,inner sep=0,minimum size=2mm,fill=black,label={[label distance=0]90:\textbf{NEL}}] at (axis cs: -1,-1.91) {};
        \node[circle,inner sep=0,minimum size=2mm,fill=black,label={[label distance=-4]-45:\textbf{PV}}] at (axis cs: 1.4,-0.986) {};
        \node[circle,inner sep=0,minimum size=2mm,fill=black,label={[label distance=0]90:\textbf{TR}}] at (axis cs: -1,0.26) {};
        \node[circle,inner sep=0,minimum size=2mm,fill=black,label={[label distance=-4]135:\textbf{LED}}] at (axis cs: 1.4,1.68) {};
        \node[circle,inner sep=0,minimum size=.1mm,fill=none,label={[label distance=0,color=blue]90:$\Tc$}] at (axis cs: -4,-1.997) {};
        \node[circle,inner sep=0,minimum size=.1mm,fill=none,label={[label distance=0,color=red]-90:$\Th$}] at (axis cs: -4,-0.065) {};

        \end{axis}
  
    \end{tikzpicture}
    \caption{Schematic of current-voltage characteristics of optoelectronic components, respectively when it is hotter (in red) or colder (in blue) than the surroundings. $U\cdot J>0$ means that electrical power is consumed.}
    \label{fig:supp_IV}
    \vspace{1.5cm}
    \centering
    \begin{tikzpicture}[charge/.style={circle,text width=0 mm,minimum size=1 mm},
                        electron/.style={charge,colfe},
                        hole/.style={charge,colfh},
                        contact/.style={rectangle,minimum width=5 mm,minimum height=2 mm},
                        econtact/.style={contact,top color=white,bottom color=blue!20!black},
                        hcontact/.style={contact,top color=red!30!black,bottom color=white},
                        arrowtherm/.style={arrow,line width=1,decorate,decoration={zigzag,amplitude=0.5mm,pre length=1 mm,post length=2 mm,segment length=1 mm}},
                        arrowcollec/.style={arrow,line width=1,decorate,decoration={zigzag,amplitude=0.5mm,pre length=8 mm,post length=8 mm,segment length=1 mm}},
                        harrowcolor/.style={coldh},
                        earrowcolor/.style={colde},
                        ray2/.style={ray,line width=2,decoration={post length=1.5 mm}}]
        \tikzmath{\x1 = 0.5;\x2 = 1;\x3 = 3.5;\x4 = 4;\x5 = 4.5;\y1 = 0.5;\y2 = 1;\dyb = 1.75;\dyth=0.8;\rth=0.8;\dyc=0.1;\dyd=5.5;\dyh=0.7;\dybelow=0.5;\fractxt=0.5;\dycontact = 0.5;}
  
        \draw[thick,draw=orange,line width=2] (-0.75 cm,\dyd cm -\dyth cm) rectangle (\x5 cm + 0.75 cm,\dyd cm + \y2 cm + \dyb cm + \dyth cm);
        \node(LED) at (\x5 cm + 0.75 cm,\dyd cm -\dyth cm) [anchor=north east] {\large\textbf{Thermoradiative cell}};
  
        \coordinate(a0) at (0,\dyd cm);
        \node at (a0) [anchor=east] {$E_V$};
        \coordinate(a1) at (\x1 cm,\dyd cm);
        \coordinate(a2) at (\x2 cm,\dyd cm + \y1 cm);
        \coordinate(a3) at (\x3 cm,\dyd cm + \y1 cm);
        \coordinate(a4) at (\x4 cm,\dyd cm + \y2 cm);
        \coordinate(a5) at (\x5 cm,\dyd cm + \y2 cm);
        \draw (a0) -- (a1) .. controls +(right:0.25cm) and +(left:0.25cm) .. (a2) -- (a3) .. controls +(right:0.25cm) and +(left:0.25cm) .. (a4) -- (a5);
  
        \coordinate(b0) at (0,\dyd cm + \dyb cm);
        \coordinate(b1) at (\x1 cm,\dyd cm + \dyb cm);
        \coordinate(b2) at (\x2 cm,\dyd cm + \y1 cm + \dyb cm);
        \coordinate(b3) at (\x3 cm,\dyd cm + \y1 cm + \dyb cm);
        \coordinate(b4) at (\x4 cm,\dyd cm + \y2 cm + \dyb cm);
        \coordinate(b5) at (\x5 cm,\dyd cm + \y2 cm + \dyb cm);
        \node at (b5) [anchor=west] {$E_C$};
        \draw (b0) -- (b1) .. controls +(right:0.25cm) and +(left:0.25cm) .. (b2) -- (b3) .. controls +(right:0.25cm) and +(left:0.25cm) .. (b4) -- (b5);
  
        \node(holeLEDa) at (0.5*\x5 cm,\dyd cm + \y1 cm - \dyth cm) [hole] {};
        \node(holeLEDb) at (0.5*\x5 cm+\rth*\dyth cm,\dyd cm + \y1 cm - \dyth cm + \rth*\dyth cm) [hole] {};
        \node(holeLEDcontact) at (\x5 cm + 0.25 cm,\dyd cm + \y2 cm + \dycontact cm) [hcontact] {};
        \draw[arrow,line width=1,harrowcolor] (holeLEDb) to (holeLEDa);
        \draw[arrow,line width=1,harrowcolor,<-] (holeLEDb) .. controls +(right:0.25*\x5 cm) .. (holeLEDcontact.north west);
        \draw[dashed,red!30!black] (holeLEDcontact.north west) to ($(holeLEDcontact.north west) - (\x5 cm,0)$);
  
        \node(electronLEDa) at (0.5*\x5 cm,\dyd cm + \y1 cm + \dyb cm + \dyth cm) [electron] {};
        \node(electronLEDb) at (0.5*\x5 cm-\rth*\dyth cm,\dyd cm + \y1 cm + \dyb cm + \dyth cm - \rth*\dyth cm) [electron] {};
        \node(electronLEDcontact) at (- 0.25 cm,\dyd cm + \dyb cm - \dycontact cm) [econtact] {};
        \draw[arrow,line width=1,earrowcolor] (electronLEDb) to (electronLEDa);
        \draw[arrow,line width=1,earrowcolor,<-] (electronLEDb) .. controls +(left:0.25*\x5 cm) .. (electronLEDcontact.south east);
        \draw[arrow,line width=1.5] (electronLEDa) to (holeLEDa);
        \draw[dashed,blue!20!black] (electronLEDcontact.south east) to ($(electronLEDcontact.south east) + (\x5 cm,0)$);
        \draw[<-,>=stealth] ($(electronLEDcontact.south east) + (\x5 cm - 5 mm,0)$) to node[anchor=east]{$\muh$} ($(holeLEDcontact.north west) + (- 5 mm,0)$);
  
        \draw[thick,draw=blue,line width=2] (-0.75 cm,-\dyth cm) rectangle (\x5 cm + 0.75 cm,\y2 cm + \dyb cm + \dyth cm);
        \node(PV) at (\x5 cm + 0.75 cm,\y2 cm + \dyb cm + \dyth cm) [anchor=south east] {\large\textbf{NEL diode}};
  
        \coordinate(c0) at (0,\y2 cm);
        \coordinate(c1) at (\x1 cm,\y2 cm);
        \coordinate(c2) at (\x2 cm,\y1 cm);
        \coordinate(c3) at (\x3 cm,\y1 cm);
        \coordinate(c4) at (\x4 cm,0);
        \coordinate(c5) at (\x5 cm,0);
        \node at (c5) [anchor=west] {$E_V$};
        \draw (c0) -- (c1) .. controls +(right:0.25cm) and +(left:0.25cm) .. (c2) -- (c3) .. controls +(right:0.25cm) and +(left:0.25cm) .. (c4) -- (c5);
  
        \coordinate(d0) at (0,\y2 cm + \dyb cm);
        \coordinate(d1) at (\x1 cm,\y2 cm + \dyb cm);
        \coordinate(d2) at (\x2 cm,\y1 cm + \dyb cm);
        \coordinate(d3) at (\x3 cm,\y1 cm + \dyb cm);
        \coordinate(d4) at (\x4 cm,\dyb cm);
        \coordinate(d5) at (\x5 cm,\dyb cm);
        \node at (d0) [anchor=east] {$E_C$};
        \draw (d0) -- (d1) .. controls +(right:0.25cm) and +(left:0.25cm) .. (d2) -- (d3) .. controls +(right:0.25cm) and +(left:0.25cm) .. (d4) -- (d5);
  
        \node(holePVa) at (0.5*\x5 cm,\y1 cm - \dyth cm) [hole] {};
        \node(holePVb) at (0.5*\x5 cm-\rth*\dyth cm,\y1 cm - \dyth cm + \rth*\dyth cm) [hole] {};
        \node(holePVcontact) at (- 0.25 cm,\y2 cm + \dycontact cm) [hcontact] {};
        \draw[arrow,line width=1,harrowcolor] (holePVa) to (holePVb);
        \draw[arrow,line width=1,harrowcolor] (holePVb) .. controls +(left:0.25*\x5 cm) .. (holePVcontact.north east);
        \draw[dashed,red!30!black] (holePVcontact.north east) to ($(holePVcontact.north east) + (\x5 cm,0)$);
  
        \node(electronPVa) at (0.5*\x5 cm,\y1 cm + \dyb cm + \dyth cm) [electron] {};
        \node(electronPVb) at (0.5*\x5 cm+\rth*\dyth cm,\y1 cm + \dyb cm + \dyth cm - \rth*\dyth cm) [electron] {};
        \node(electronPVcontact) at (\x5 cm + 0.25 cm,\dyb cm - \dycontact cm) [econtact] {};
        \draw[arrow,line width=1,earrowcolor] (electronPVa) to (electronPVb);
        \draw[arrow,line width=1,earrowcolor] (electronPVb) .. controls +(right:0.25*\x5 cm) .. (electronPVcontact.south west);
        \draw[arrow,line width=1.5] (holePVa) to (electronPVa);
        \draw[dashed,blue!20!black] (electronPVcontact.south west) to ($(electronPVcontact.south west) - (\x5 cm,0)$);
        \draw[->,>=stealth] ($(holePVcontact.north east) + (\x5 cm - 5 mm,0)$) to node[anchor=east]{$\muc$} ($(electronPVcontact.south west) + (- 5 mm,0)$);
  
        \node(heatsource) at (\x5 cm + 0.75 cm,\dyd cm + \y2 cm + \dyb cm + \dyth cm + \dyh cm) [anchor=south east,minimum width = \x5 cm + 1.5 cm,minimum height = 0.5cm,color_heatsource] {\large Heat source};
        \node(heatsink)   at (-0.75 cm,-\dyth cm - \dyh cm) [anchor=north west,minimum width = \x5 cm + 1.5 cm,minimum height = 0.5cm,color_heatsink] {\large Heat sink};
  
        \draw[arrow,<-] ($0.5*(electronLEDa.south west)+0.5*(electronLEDb.north east) + (-1 mm,0 mm)$) to ($(heatsource.south west)+(0.75 cm + 0.5*\x5 cm - 0.5*\rth*\dyth cm,0) + (-1 mm,0)$) node[right,yshift=-\fractxt*\dyh cm] {\large$\hphantom{0}\mathsub{q}{source}$} ;
        \draw[arrow,<-] ($(electronLEDb)-(0.4 cm,-0.5 mm)$) to ($(heatsource.south west)+(0.75 cm + 0.5*\x5 cm - \rth*\dyth cm,0)-(0.4 cm,0)$);
        \draw[arrow] ($0.5*(holePVa.south east)+0.5*(holePVb.north west) + (-1 mm,-1 mm)$) to ($(heatsink.north west)+(0.75 cm + 0.5*\x5 cm - 0.5*\rth*\dyth cm,0) + (-1 mm,0)$) node[right,yshift=\fractxt*\dyh cm] {\large$\hphantom{0}\mathsub{q}{sink}$};
        \draw[arrow] ($(holePVb)-(0.4 cm,1 mm)$) to ($(heatsink.north west)+(0.75 cm + 0.5*\x5 cm - \rth*\dyth cm,0)-(0.4 cm,0)$);
        \draw[ray2] ($0.5*(electronLEDb)+0.5*(holeLEDb)$) .. controls +(left:0.4*\x5 cm) .. (0.1*\x5 cm,\dyd cm -\dyth cm) to node[left,yshift=.0cm] {\large$q,\dot{N}\hphantom{0}$} (0.1*\x5 cm,\y2 cm + \dyb cm + \dyth cm) .. controls (0.1*\x5 cm,\y1 cm + 0.5*\dyb cm) .. ($0.5*(electronPVa)+0.5*(holePVa)$);
  
        \draw[arrow] (\x5 cm + .75 cm,\dyd cm + 0.5*\y2 cm + 0.5*\dyb cm) to (\x5 cm + 1.25 cm,\dyd cm + 0.5*\y2 cm + 0.5*\dyb cm) node[right] {\large$P_\mathrm{h}=\muh\dot{N}$};
        \draw[arrow,<-] (\x5 cm + .75 cm,0.5*\y2 cm + 0.5*\dyb cm) to (\x5 cm + 1.25 cm,0.5*\y2 cm + 0.5*\dyb cm) node[right] {\large$P_\mathrm{c}=\muc\dot{N}$};
  
    \end{tikzpicture}
    \caption{Schematic of the energy flow through a thermoradiative-negative electroluminescent (TRNEL) heat engine at the radiative limit, along with the energy diagram of the two components.}
    \label{fig:supp_SchemaEnergy}
\end{figure}

We provide in Fig. \ref{fig:supp_IV} the current-voltage characteristic of the various optoelectronic components considered for dual radiative engines. Each component operates in a separate quadrant, two of them producing electrical power (TR and PV cells) while the two others consume power (LEDs and NEL diodes). 

As an example, we also present in Fig. \ref{fig:supp_SchemaEnergy} the energy diagrams of a TR cell and a NEL diode used in a TRNEL device, along with the various flows of energy exchanged. The negative bias applied to both components allows the charge concentration in the respective bands to be reduced, resulting in a decrease in emission. Because the TR cell is maintained at high temperature, it will nonetheless emit towards the cold NEL diode, generating electrical power. The low emission of the NEL diode further reduces the generation of electron-hole pairs in the TR cell that would otherwise repopulate the bands.

Additional information about the respective operating principle and energy diagram of the individual components employed in dual radiative heat engines can be found in \cite{Tervo2018}.

\begin{figure}[!htp]
    \centering
    \begin{subfigure}{0.48\textwidth}
        \centering
        \includegraphics{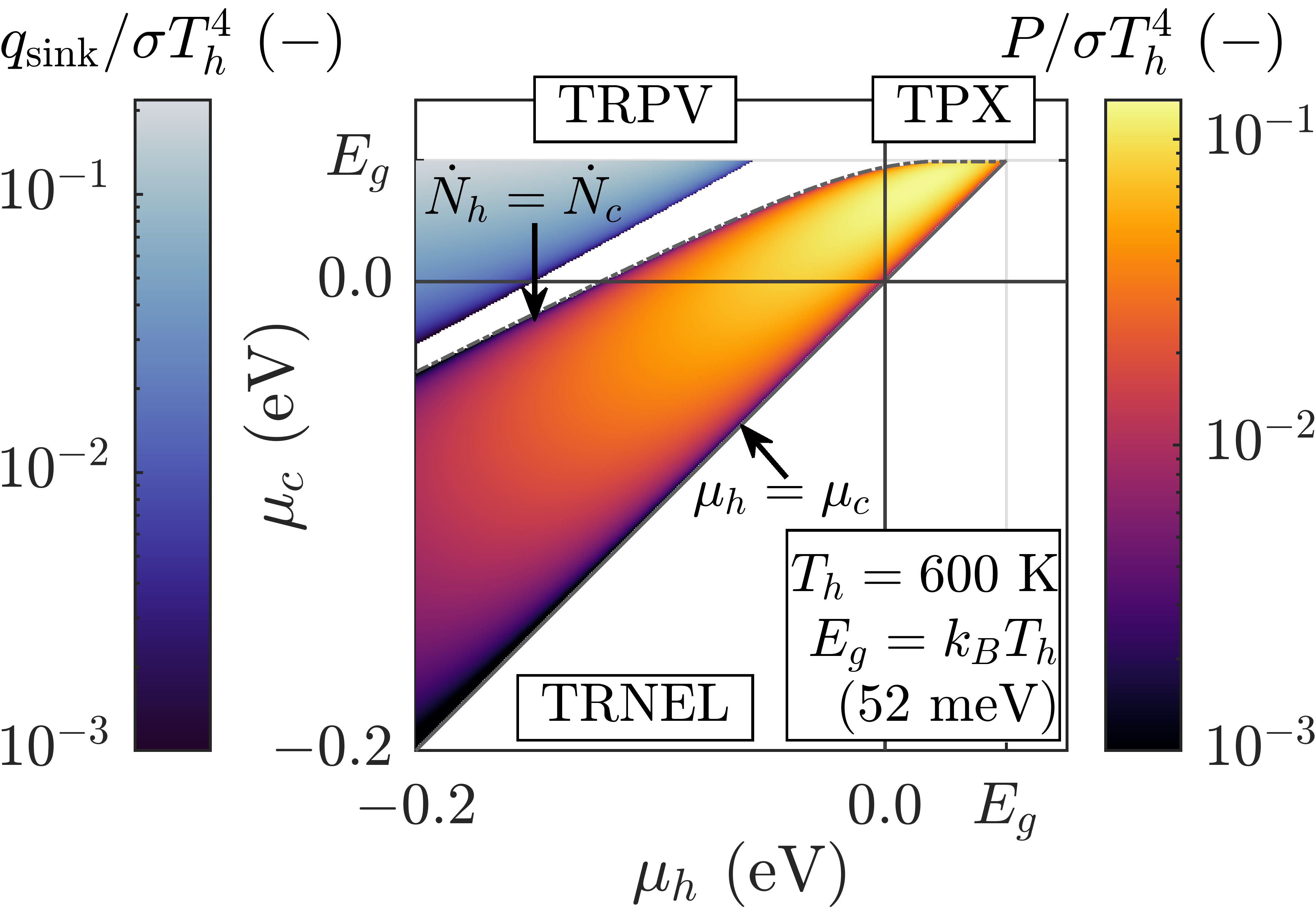}
        \caption{}
        \label{fig:supp_mumu}
    \end{subfigure}
    \begin{subfigure}{0.48\textwidth}
        \centering
        \includegraphics{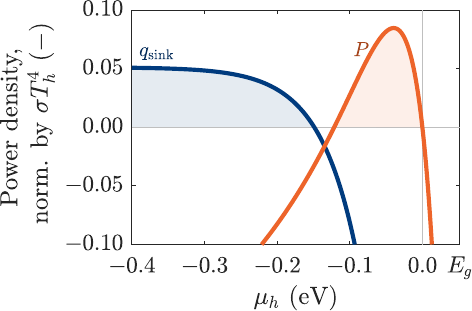}
        \caption{}
        \label{fig:supp_muhP}
    \end{subfigure}
    
    \caption{(a) Performance of dual radiative systems operating as heat engines or heat pumps, at the radiative limit and for $\Eg=\kb \Th=52$ meV. (b) Zoom on the $\muc=0$ scenario.}
    \label{fig:supp_mu}
\end{figure}

\section{Performance of dual engines for low bandgaps}\label{app:lowEg}

We provide in Fig. \ref{fig:supp_mu} the variation of electrical power output and cooling power as a function of both chemical potentials, this time for $\Eg=\kb\Th$. In comparison to the results obtained for larger bandgaps, the gap between the heat engine and heat pump operating regions is larger. The change is particularly visible in the TPX quadrant, since then $\Eg-\mu_i$ becomes lower than $\kb T_i$. In this quadrant, the device is almost always capable of operating as a heat engine as long as $\muc\ge \muh$. Consequently, TPX devices are not able to operate as heat pumps for such low bandgaps.

\section{Impact of heat source temperature on $\eta-P$ characteristics}\label{app:impactT}

\begin{figure*}[t]
    \hfill
    \begin{minipage}{.48\textwidth}
        \centering
        \includegraphics{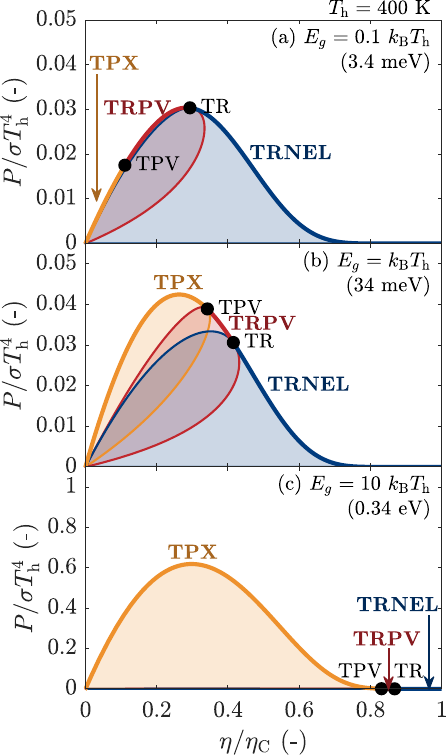}
        \caption{$\eta-P$ plots obtained for dual radiative engines at the radiative limit, for $\Th=400$ K and for various bandgaps.}
        \label{fig:supp_400K}
    \end{minipage}
    \hfill
    \begin{minipage}{.48\textwidth}
        \centering
        \includegraphics{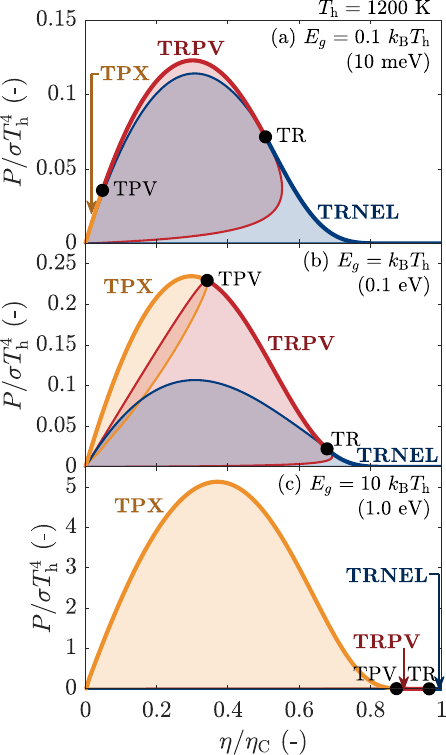}
        \caption{$\eta-P$ plots obtained for dual radiative engines at the radiative limit, for $\Th=1200$ K and for various bandgaps.}
        \label{fig:supp_1200K}
    \end{minipage}
    \hfill
\end{figure*}

Figures \ref{fig:supp_400K} and \ref{fig:supp_1200K} show $\eta-P$ plots obtained at the radiative limit, for heat source temperatures of 400 K and 1200 K respectively. The conclusion drawn for $\Th=600$ K remains valid in these scenarios. We can still notice two slight differences. First, the normalised power output achieved becomes larger as $\Th$ increases. For $\Eg=10\kb\Th$, for instance, $\mathsub{P}{max}/\sigma\Th^4$ equals approximately 0.6 for $\Th=400$ K, but rises to 2.4 for $\Th=600$ K and exceeds 5 for $\Th=1200$ K. Second, interest in TRPV devices rises with temperature, more and more of the total envelope corresponding to that of the TRPV device if the bandgap stays moderate. For $\Th=1200$ K and $\Eg=\kb\Th\approx 0.1$ eV, TRPV gives access to interesting trade-offs between power and efficiency. However, note that such high temperatures can hardly be withstood by optoelectronic components, which therefore limits interest in TRPV.

\section{Variation of maximum power and related efficiency with bandgap}\label{app:impactEg}

\begin{figure}[!ht]
    \centering
    \includegraphics{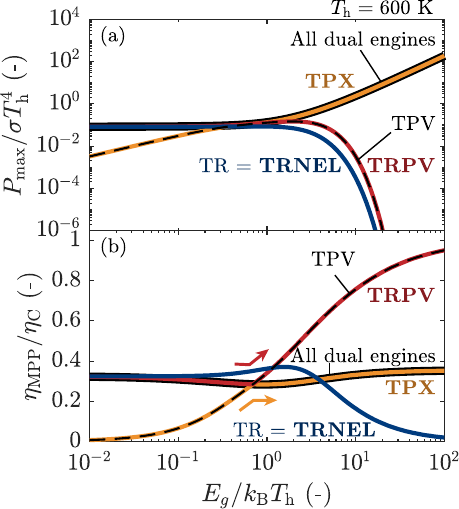}
    \caption{Variation of (a) the maximum power, (b) the efficiency at maximum power, for the different radiative engines considering $\Th=600$ K.}
    \label{fig:supp_Peta_fEg}
\end{figure}

We provide in Fig. \ref{fig:supp_Peta_fEg} the variations of the maximum power and related efficiency as a function of the bandgap, for the complete dual engine (thick black line) and for each individual engine. For a heat source temperature of 600 K, the dual-engine MPP moves from the TRPV quadrant for low bandgaps to the TPX quadrant for higher bandgaps. In this case, the transition between the two quadrants occurs around $\Eg/\kb\Th=0.7$. Dual engines become especially attractive when $\Eg>\kb\Th\approx 0.05$ eV, their power output increasing with $\Eg$ and being already 50\% larger than that of single engines for $\Eg=1.9\kb\Th$. In practice, this condition is satisfied by any realistic bandgap at the temperature considered.

Note how, when TPX or TRPV engines do not maximise the power output, they operate as TPV devices. This is what causes the sudden change in slope indicated by arrows in panel (b): around $\Eg/\kb\Th=0.7$, there is for both devices an abrupt change in the direction of displacement of the MPP in $(\muh,\muc)$ coordinates, from that of the complete dual engine to that of a TPV device (or vice versa).

One can also observe that optimising TRNEL devices for electrical power production simply means operating as a TR device. Moreover, in the limit of zero bandgap, TR operation becomes optimal: this is because the TRNEL quadrant is the only one available since $\mu<\Eg\rightarrow 0$.

\section{Dependency of the optimum bandgap on the quantum efficiency}\label{app:optEg}

\begin{figure}[t]
    \centering
    \includegraphics{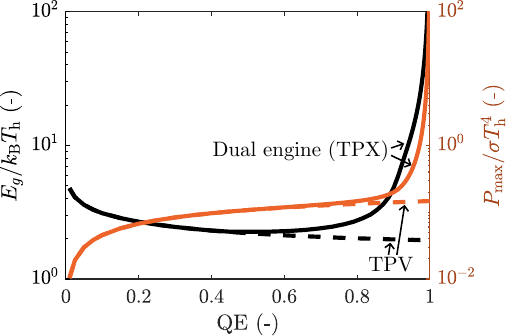}
    \caption{Variation of the optimum bandgap (in terms of power output) with quantum efficiency, considering $\Th=600$ K.}
    \label{fig:optEg}
\end{figure}

We show in Fig. \ref{fig:optEg} how the bandgap leading to maximum power varies with the quantum efficiency, along with the variation of the maximum power itself. First, as mentioned in Section \ref{sec:etaP}, the optimum bandgap diverges when QE approaches 1, that is, when the engine approaches the radiative limit. Close to the divergence, the optimum bandgap and the maximum power have been observed to vary as $1/(1-\mathrm{QE})$ and $1/(1-\mathrm{QE})^2$, respectively. In Section \ref{sec:nonrad}, we highlighted that the power output enhancement obtained by using dual engines (in the TPX regime) over TPV devices was only significant for large enough QEs. This can clearly be observed here, with the optimum bandgap and the maximum power of dual engines deviating from those of TPV for QEs above 0.5 and 0.75, respectively. This deviation becomes especially prominent for $\mathrm{QE}\ge 0.9$. As opposed to TPV, the optimum bandgap of dual engines actually varies non-monotonically with QE: it reaches a minimum close to $\mathrm{QE}= 0.5$, and then starts increasing.

\section{Determination of the maximum power and related efficiency for broadband radiation}\label{app:analyticBroad}

The goal of this section is to provide certain details allowing to determine analytical expressions of the maximum power and related efficiency, in the limit of $\Eg\rightarrow\infty$. Using the quantity $x_{i}=(\Eg-\mu_i)/\kb T_i$, we obtain that $\muc-\muh=x_{h}\kb\Th-x_{c}\kb\Tc$. Defining $X_i$ as $\exp(-x_i)$, the power output can then be expressed as:
\begin{align}
    \begin{split}
        P=&\frac{\Eg^2\kb^2\Th^2}{4\pi^2c^2\hbar^3}\left(\ln(\hot{X})-\frac{\Tc}{\Th}\ln(\cold{X})\right)\\
          &\times\left(\ln(1-\hot{X})-\frac{\Tc}{\Th}\ln(1-\cold{X})\right),
    \end{split}
\end{align}
since $\polylog_1(x)=\ln(1-x)$. First, we must verify whether the maximum power is reached inside the $(\hot{X},\cold{X})$ domain (i.e. for $0<X_i<1$) or at the boundary. For instance, if $\hot{X}$ goes to 0, it gives
\begin{equation}
    P\sim-\frac{\Eg^2\kb ^2\Th ^2}{4\pi^2c^2\hbar^3}\ln(\hot{X})\ln(1-\cold{X})\frac{\Tc}{\Th}\rightarrow-\infty,
\end{equation}
and the maximum power is therefore not reached at this boundary (the power being negative). A similar argument can be made at the three other boundaries to ensure that the maximum power point is indeed located inside the domain. We should then find the pair $(\hot{X},\cold{X})$ which makes both partial derivatives equal to zero. If the pair found is unique, it is the maximum power point since $P$ goes to $-\infty$ at the boundaries (which means that a maximum power point must exist).

Setting both partial derivatives of $P$ with respect to $X_i$ to zero, we get
\begin{equation}\label{eq:appThermoAn_partDer}
    \frac{\ln(1-\hot{X})-\ln(1-\cold{X})\frac{\Tc }{\Th }}{\ln(\hot{X})-\ln(\cold{X})\frac{\Tc }{\Th }}=\frac{\hot{X}}{1-\hot{X}}=\frac{\cold{X}}{1-\cold{X}}.
\end{equation}
Note that this holds only if $\ln(\hot{X})-\ln(\cold{X})\frac{\Tc }{\Th }$ is non-zero at the maximum power point. If the former term was equal to zero, then $\ln(1-\hot{X})-\ln(1-\cold{X})\frac{\Tc }{\Th }$ should also be zero to satisfy that partial derivatives are equal to zero, which would lead to $1-\cold{X}^{\Tc /\Th }=(1-\cold{X})^{\Tc /\Th }$. This equation being satisfied only for $\cold{X}$ equal to 0 or 1 (which are not inside the domain), $\ln(\hot{X})-\ln(\cold{X})\frac{\Tc }{\Th }$ is therefore non-zero at the maximum power point.

Combining the two expressions given in Eq. \eqref{eq:appThermoAn_partDer}, we easily obtain that $\hot{X}=\cold{X}$. Writing this common value as $X$, we obtain
\begin{equation}
    P=\frac{1}{\hbar}\left(\frac{\Eg\kb(\Th-\Tc)}{2\pi c\hbar}\right)^2\ln(X)\ln(1-X).
\end{equation}
A quick study of the function $\ln(X)\ln(1-X)$ immediately reveals that it reaches a maximum at $(\hot{X}=1/2,\cold{X}=1/2)$, which therefore corresponds to the maximum power point.

\begin{figure}
    \centering
    \includegraphics{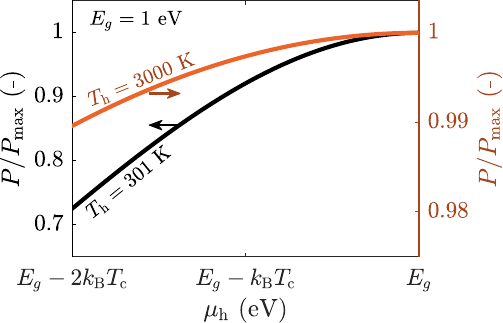}
    \caption{Variation of the power output with the LED chemical potential considering quasi-monochromatic radiation exchanged, for $\Eg=1$ eV and for two different heat source temperatures. In this case, only the TPX quadrant is considered.}
    \label{fig:supp_mumonochrom}
\end{figure}

To derive a closed-form expression for the efficiency at maximum power, we use that $x_{h\mathrm{,MPP}}=x_{c\mathrm{,MPP}}=\ln(2)$ to obtain $(\mu_{c\mathrm{,MPP}}-\mu_{h\mathrm{,MPP}})/(\Eg-\mu_{h\mathrm{,MPP}})=\etaC$. Dividing both the numerator and the denominator of Eq. \eqref{eq:efficiency} of the main text by $(\Nh-\Nc)(\Eg-\muh)$ yields
\begin{equation}
    \mathsub{\eta}{MPP}=\frac{\etaC}{1+\left(\frac{1}{\rho}-1\right)\frac{1}{\ln(2)}\frac{\Eg}{\kb \Th}},
\end{equation}
where $\rho=\Eg(\Nh-\Nc)/(\hot{q}-\cold{q})$ corresponds to the fraction of radiative energy being useful to optoelectronic conversion. To express this, both $\polylog_1$ and $\polylog_2$ terms are necessary. One obtains
\begin{equation}
    \left(\frac{1}{\rho}-1\right)\frac{1}{\ln(2)}\frac{\Eg}{\kb\Th}\underset{\Eg\rightarrow\infty}{\sim}(2-\etaC)\frac{1}{\ln(2)}\frac{\polylog_2(1/2)}{\polylog_1(1/2)}.
\end{equation}
The polylogarithmic terms having closed-form expressions for $x=1/2$, the efficiency at maximum power obtained as $\Eg\rightarrow\infty$ is
\begin{equation}
    \eta_{\mathrm{MPP}}^{\Eg\gg\kb\Th}=\frac{\etaC}{1+(2-\etaC)\chi},
\end{equation}
$\chi$ being a constant defined in the main paper.

\section{Determination of the efficiency at maximum power for quasi-monochromatic radiation}\label{app:analyticNarrow}

To derive an expression for the efficiency at maximum power in the case of quasi-monochromatic radiation, we use that $\Eg-\mu_i\ll \kb T_i$. To verify that this is indeed correct, we show in Fig. \ref{fig:supp_mumonochrom} the variation of power output with $\muh$ for $\Eg=1$ eV, considering only the TPX quadrant. The power output associated to each $\muh$ corresponds to the maximum achievable with the whole range of $\muc$ available. We observe that $\Eg-\mu_{h\mathrm{,MPP}}\ll \kb\Tc$: since $\Th>\Tc$, therefore $\Eg-\mu_{h\mathrm{,MPP}}\ll \kb\Th$. Moreover, $\muc\ge \muh$, and thus $\Eg-\mu_{c\mathrm{,MPP}}\ll \kb\Tc$.

Because $\Eg-\mu_i\ll \kb T_i$, the Bose-Einstein distributions can be simplified using that $[\exp(x)-1]^{-1}\sim x^{-1}$ around 0. Setting to zero any of the two partial derivatives of $P$ with respect to $\mu_i$ then leads to
\begin{equation}
    \eta_{\mathrm{MPP}}^{\delta E\ll\kb\Tc}=1-\sqrt{\frac{\Tc}{\Th}}.
\end{equation}

\vspace{1.5cm}

\bibliography{bibliography}

\begin{thebibliography}{40}%
\makeatletter
\providecommand \@ifxundefined [1]{%
 \@ifx{#1\undefined}
}%
\providecommand \@ifnum [1]{%
 \ifnum #1\expandafter \@firstoftwo
 \else \expandafter \@secondoftwo
 \fi
}%
\providecommand \@ifx [1]{%
 \ifx #1\expandafter \@firstoftwo
 \else \expandafter \@secondoftwo
 \fi
}%
\providecommand \natexlab [1]{#1}%
\providecommand \enquote  [1]{``#1''}%
\providecommand \bibnamefont  [1]{#1}%
\providecommand \bibfnamefont [1]{#1}%
\providecommand \citenamefont [1]{#1}%
\providecommand \href@noop [0]{\@secondoftwo}%
\providecommand \href [0]{\begingroup \@sanitize@url \@href}%
\providecommand \@href[1]{\@@startlink{#1}\@@href}%
\providecommand \@@href[1]{\endgroup#1\@@endlink}%
\providecommand \@sanitize@url [0]{\catcode `\\12\catcode `\$12\catcode `\&12\catcode `\#12\catcode `\^12\catcode `\_12\catcode `\%12\relax}%
\providecommand \@@startlink[1]{}%
\providecommand \@@endlink[0]{}%
\providecommand \url  [0]{\begingroup\@sanitize@url \@url }%
\providecommand \@url [1]{\endgroup\@href {#1}{\urlprefix }}%
\providecommand \urlprefix  [0]{URL }%
\providecommand \Eprint [0]{\href }%
\providecommand \doibase [0]{https://doi.org/}%
\providecommand \selectlanguage [0]{\@gobble}%
\providecommand \bibinfo  [0]{\@secondoftwo}%
\providecommand \bibfield  [0]{\@secondoftwo}%
\providecommand \translation [1]{[#1]}%
\providecommand \BibitemOpen [0]{}%
\providecommand \bibitemStop [0]{}%
\providecommand \bibitemNoStop [0]{.\EOS\space}%
\providecommand \EOS [0]{\spacefactor3000\relax}%
\providecommand \BibitemShut  [1]{\csname bibitem#1\endcsname}%
\let\auto@bib@innerbib\@empty
\bibitem [{\citenamefont {Melnick}\ and\ \citenamefont {Kaviany}(2019)}]{Melnick2019}%
  \BibitemOpen
  \bibfield  {author} {\bibinfo {author} {\bibfnamefont {C.}~\bibnamefont {Melnick}}\ and\ \bibinfo {author} {\bibfnamefont {M.}~\bibnamefont {Kaviany}},\ }\bibfield  {title} {\bibinfo {title} {From thermoelectricity to phonoelectricity},\ }\href {https://doi.org/10.1063/1.5031425} {\bibfield  {journal} {\bibinfo  {journal} {Applied Physics Reviews}\ }\textbf {\bibinfo {volume} {6}},\ \bibinfo {pages} {021305} (\bibinfo {year} {2019})}\BibitemShut {NoStop}%
\bibitem [{\citenamefont {Shi}\ \emph {et~al.}(2020)\citenamefont {Shi}, \citenamefont {Zou},\ and\ \citenamefont {Chen}}]{Shi2020}%
  \BibitemOpen
  \bibfield  {author} {\bibinfo {author} {\bibfnamefont {X.-L.}\ \bibnamefont {Shi}}, \bibinfo {author} {\bibfnamefont {J.}~\bibnamefont {Zou}},\ and\ \bibinfo {author} {\bibfnamefont {Z.-G.}\ \bibnamefont {Chen}},\ }\bibfield  {title} {\bibinfo {title} {Advanced {{Thermoelectric Design}}: {{From Materials}} and {{Structures}} to {{Devices}}},\ }\href {https://doi.org/10.1021/acs.chemrev.0c00026} {\bibfield  {journal} {\bibinfo  {journal} {Chemical Reviews}\ }\textbf {\bibinfo {volume} {120}},\ \bibinfo {pages} {7399} (\bibinfo {year} {2020})}\BibitemShut {NoStop}%
\bibitem [{\citenamefont {Abdul~Khalid}\ \emph {et~al.}(2016)\citenamefont {Abdul~Khalid}, \citenamefont {Leong},\ and\ \citenamefont {Mohamed}}]{AbdulKhalid2016}%
  \BibitemOpen
  \bibfield  {author} {\bibinfo {author} {\bibfnamefont {K.~A.}\ \bibnamefont {Abdul~Khalid}}, \bibinfo {author} {\bibfnamefont {T.~J.}\ \bibnamefont {Leong}},\ and\ \bibinfo {author} {\bibfnamefont {K.}~\bibnamefont {Mohamed}},\ }\bibfield  {title} {\bibinfo {title} {Review on {{Thermionic Energy Converters}}},\ }\href {https://doi.org/10.1109/TED.2016.2556751} {\bibfield  {journal} {\bibinfo  {journal} {IEEE Transactions on Electron Devices}\ }\textbf {\bibinfo {volume} {63}},\ \bibinfo {pages} {2231} (\bibinfo {year} {2016})}\BibitemShut {NoStop}%
\bibitem [{\citenamefont {Burger}\ \emph {et~al.}(2020)\citenamefont {Burger}, \citenamefont {Sempere}, \citenamefont {{Roy-Layinde}},\ and\ \citenamefont {Lenert}}]{Burger2020}%
  \BibitemOpen
  \bibfield  {author} {\bibinfo {author} {\bibfnamefont {T.}~\bibnamefont {Burger}}, \bibinfo {author} {\bibfnamefont {C.}~\bibnamefont {Sempere}}, \bibinfo {author} {\bibfnamefont {B.}~\bibnamefont {{Roy-Layinde}}},\ and\ \bibinfo {author} {\bibfnamefont {A.}~\bibnamefont {Lenert}},\ }\bibfield  {title} {\bibinfo {title} {Present {{Efficiencies}} and {{Future Opportunities}} in {{Thermophotovoltaics}}},\ }\href {https://doi.org/10.1016/j.joule.2020.06.021} {\bibfield  {journal} {\bibinfo  {journal} {Joule}\ }\textbf {\bibinfo {volume} {4}},\ \bibinfo {pages} {1660} (\bibinfo {year} {2020})}\BibitemShut {NoStop}%
\bibitem [{\citenamefont {LaPotin}\ \emph {et~al.}(2022)\citenamefont {LaPotin}, \citenamefont {Schulte}, \citenamefont {Steiner}, \citenamefont {Buznitsky}, \citenamefont {Kelsall}, \citenamefont {Friedman}, \citenamefont {Tervo}, \citenamefont {France}, \citenamefont {Young}, \citenamefont {Rohskopf}, \citenamefont {Verma}, \citenamefont {Wang},\ and\ \citenamefont {Henry}}]{LaPotin2022}%
  \BibitemOpen
  \bibfield  {author} {\bibinfo {author} {\bibfnamefont {A.}~\bibnamefont {LaPotin}}, \bibinfo {author} {\bibfnamefont {K.~L.}\ \bibnamefont {Schulte}}, \bibinfo {author} {\bibfnamefont {M.~A.}\ \bibnamefont {Steiner}}, \bibinfo {author} {\bibfnamefont {K.}~\bibnamefont {Buznitsky}}, \bibinfo {author} {\bibfnamefont {C.~C.}\ \bibnamefont {Kelsall}}, \bibinfo {author} {\bibfnamefont {D.~J.}\ \bibnamefont {Friedman}}, \bibinfo {author} {\bibfnamefont {E.~J.}\ \bibnamefont {Tervo}}, \bibinfo {author} {\bibfnamefont {R.~M.}\ \bibnamefont {France}}, \bibinfo {author} {\bibfnamefont {M.~R.}\ \bibnamefont {Young}}, \bibinfo {author} {\bibfnamefont {A.}~\bibnamefont {Rohskopf}}, \bibinfo {author} {\bibfnamefont {S.}~\bibnamefont {Verma}}, \bibinfo {author} {\bibfnamefont {E.~N.}\ \bibnamefont {Wang}},\ and\ \bibinfo {author} {\bibfnamefont {A.}~\bibnamefont {Henry}},\ }\bibfield  {title} {\bibinfo {title} {Thermophotovoltaic efficiency of 40\%},\ }\href {https://doi.org/10.1038/s41586-022-04473-y} {\bibfield  {journal} {\bibinfo  {journal} {Nature}\ }\textbf {\bibinfo {volume} {604}},\ \bibinfo {pages} {287} (\bibinfo {year} {2022})}\BibitemShut {NoStop}%
\bibitem [{\citenamefont {Tervo}\ \emph {et~al.}(2022)\citenamefont {Tervo}, \citenamefont {France}, \citenamefont {Friedman}, \citenamefont {Arulanandam}, \citenamefont {King}, \citenamefont {Narayan}, \citenamefont {Luciano}, \citenamefont {Nizamian}, \citenamefont {Johnson}, \citenamefont {Young}, \citenamefont {Kuritzky}, \citenamefont {Perl}, \citenamefont {Limpinsel}, \citenamefont {Kayes}, \citenamefont {Ponec}, \citenamefont {Bierman}, \citenamefont {Briggs},\ and\ \citenamefont {Steiner}}]{Tervo2022}%
  \BibitemOpen
  \bibfield  {author} {\bibinfo {author} {\bibfnamefont {E.~J.}\ \bibnamefont {Tervo}}, \bibinfo {author} {\bibfnamefont {R.~M.}\ \bibnamefont {France}}, \bibinfo {author} {\bibfnamefont {D.~J.}\ \bibnamefont {Friedman}}, \bibinfo {author} {\bibfnamefont {M.~K.}\ \bibnamefont {Arulanandam}}, \bibinfo {author} {\bibfnamefont {R.~R.}\ \bibnamefont {King}}, \bibinfo {author} {\bibfnamefont {T.~C.}\ \bibnamefont {Narayan}}, \bibinfo {author} {\bibfnamefont {C.}~\bibnamefont {Luciano}}, \bibinfo {author} {\bibfnamefont {D.~P.}\ \bibnamefont {Nizamian}}, \bibinfo {author} {\bibfnamefont {B.~A.}\ \bibnamefont {Johnson}}, \bibinfo {author} {\bibfnamefont {A.~R.}\ \bibnamefont {Young}}, \bibinfo {author} {\bibfnamefont {L.~Y.}\ \bibnamefont {Kuritzky}}, \bibinfo {author} {\bibfnamefont {E.~E.}\ \bibnamefont {Perl}}, \bibinfo {author} {\bibfnamefont {M.}~\bibnamefont {Limpinsel}}, \bibinfo {author} {\bibfnamefont {B.~M.}\ \bibnamefont {Kayes}}, \bibinfo {author} {\bibfnamefont {A.~J.}\ \bibnamefont {Ponec}}, \bibinfo {author} {\bibfnamefont {D.~M.}\ \bibnamefont {Bierman}}, \bibinfo {author} {\bibfnamefont {J.~A.}\ \bibnamefont {Briggs}},\ and\ \bibinfo {author} {\bibfnamefont {M.~A.}\ \bibnamefont {Steiner}},\ }\bibfield  {title} {\bibinfo {title} {Efficient and scalable {{GaInAs}} thermophotovoltaic devices},\ }\href {https://doi.org/10.1016/j.joule.2022.10.002} {\bibfield  {journal} {\bibinfo  {journal} {Joule}\ }\textbf {\bibinfo {volume} {6}},\ \bibinfo {pages} {2566} (\bibinfo {year} {2022})}\BibitemShut {NoStop}%
\bibitem [{\citenamefont {Giteau}\ \emph {et~al.}(2023)\citenamefont {Giteau}, \citenamefont {Picardi},\ and\ \citenamefont {Papadakis}}]{Giteau2023}%
  \BibitemOpen
  \bibfield  {author} {\bibinfo {author} {\bibfnamefont {M.}~\bibnamefont {Giteau}}, \bibinfo {author} {\bibfnamefont {M.~F.}\ \bibnamefont {Picardi}},\ and\ \bibinfo {author} {\bibfnamefont {G.~T.}\ \bibnamefont {Papadakis}},\ }\bibfield  {title} {\bibinfo {title} {Thermodynamic performance bounds for radiative heat engines},\ }\href {https://doi.org/10.1103/PhysRevApplied.20.L061003} {\bibfield  {journal} {\bibinfo  {journal} {Physical Review Applied}\ }\textbf {\bibinfo {volume} {20}},\ \bibinfo {pages} {L061003} (\bibinfo {year} {2023})}\BibitemShut {NoStop}%
\bibitem [{\citenamefont {Datas}\ \emph {et~al.}(2022)\citenamefont {Datas}, \citenamefont {{L{\'o}pez-Ceballos}}, \citenamefont {L{\'o}pez}, \citenamefont {Ramos},\ and\ \citenamefont {{del Ca{\~n}izo}}}]{Datas2022}%
  \BibitemOpen
  \bibfield  {author} {\bibinfo {author} {\bibfnamefont {A.}~\bibnamefont {Datas}}, \bibinfo {author} {\bibfnamefont {A.}~\bibnamefont {{L{\'o}pez-Ceballos}}}, \bibinfo {author} {\bibfnamefont {E.}~\bibnamefont {L{\'o}pez}}, \bibinfo {author} {\bibfnamefont {A.}~\bibnamefont {Ramos}},\ and\ \bibinfo {author} {\bibfnamefont {C.}~\bibnamefont {{del Ca{\~n}izo}}},\ }\bibfield  {title} {\bibinfo {title} {Latent heat thermophotovoltaic batteries},\ }\href {https://doi.org/10.1016/j.joule.2022.01.010} {\bibfield  {journal} {\bibinfo  {journal} {Joule}\ }\textbf {\bibinfo {volume} {6}},\ \bibinfo {pages} {418} (\bibinfo {year} {2022})}\BibitemShut {NoStop}%
\bibitem [{\citenamefont {Strandberg}(2015)}]{Strandberg2015}%
  \BibitemOpen
  \bibfield  {author} {\bibinfo {author} {\bibfnamefont {R.}~\bibnamefont {Strandberg}},\ }\bibfield  {title} {\bibinfo {title} {Theoretical efficiency limits for thermoradiative energy conversion},\ }\href {https://doi.org/10.1063/1.4907392} {\bibfield  {journal} {\bibinfo  {journal} {Journal of Applied Physics}\ }\textbf {\bibinfo {volume} {117}},\ \bibinfo {pages} {055105} (\bibinfo {year} {2015})}\BibitemShut {NoStop}%
\bibitem [{\citenamefont {Pusch}\ \emph {et~al.}(2019)\citenamefont {Pusch}, \citenamefont {Gordon}, \citenamefont {Mellor}, \citenamefont {Krich},\ and\ \citenamefont {{Ekins-Daukes}}}]{Pusch2019}%
  \BibitemOpen
  \bibfield  {author} {\bibinfo {author} {\bibfnamefont {A.}~\bibnamefont {Pusch}}, \bibinfo {author} {\bibfnamefont {J.~M.}\ \bibnamefont {Gordon}}, \bibinfo {author} {\bibfnamefont {A.}~\bibnamefont {Mellor}}, \bibinfo {author} {\bibfnamefont {J.~J.}\ \bibnamefont {Krich}},\ and\ \bibinfo {author} {\bibfnamefont {N.~J.}\ \bibnamefont {{Ekins-Daukes}}},\ }\bibfield  {title} {\bibinfo {title} {Fundamental {{Efficiency Bounds}} for the {{Conversion}} of a {{Radiative Heat Engine}}'s {{Own Emission}} into {{Work}}},\ }\href {https://doi.org/10.1103/PhysRevApplied.12.064018} {\bibfield  {journal} {\bibinfo  {journal} {Physical Review Applied}\ }\textbf {\bibinfo {volume} {12}},\ \bibinfo {pages} {064018} (\bibinfo {year} {2019})}\BibitemShut {NoStop}%
\bibitem [{\citenamefont {Santhanam}\ and\ \citenamefont {Fan}(2016)}]{Santhanam2016}%
  \BibitemOpen
  \bibfield  {author} {\bibinfo {author} {\bibfnamefont {P.}~\bibnamefont {Santhanam}}\ and\ \bibinfo {author} {\bibfnamefont {S.}~\bibnamefont {Fan}},\ }\bibfield  {title} {\bibinfo {title} {Thermal-to-electrical energy conversion by diodes under negative illumination},\ }\href {https://doi.org/10.1103/PhysRevB.93.161410} {\bibfield  {journal} {\bibinfo  {journal} {Physical Review B}\ }\textbf {\bibinfo {volume} {93}},\ \bibinfo {pages} {161410} (\bibinfo {year} {2016})}\BibitemShut {NoStop}%
\bibitem [{\citenamefont {Tervo}\ \emph {et~al.}(2018)\citenamefont {Tervo}, \citenamefont {Bagherisereshki},\ and\ \citenamefont {Zhang}}]{Tervo2018}%
  \BibitemOpen
  \bibfield  {author} {\bibinfo {author} {\bibfnamefont {E.}~\bibnamefont {Tervo}}, \bibinfo {author} {\bibfnamefont {E.}~\bibnamefont {Bagherisereshki}},\ and\ \bibinfo {author} {\bibfnamefont {Z.}~\bibnamefont {Zhang}},\ }\bibfield  {title} {\bibinfo {title} {Near-field radiative thermoelectric energy converters: A review},\ }\href {https://doi.org/10.1007/s11708-017-0517-z} {\bibfield  {journal} {\bibinfo  {journal} {Frontiers in Energy}\ }\textbf {\bibinfo {volume} {12}},\ \bibinfo {pages} {5} (\bibinfo {year} {2018})}\BibitemShut {NoStop}%
\bibitem [{\citenamefont {Liao}\ \emph {et~al.}(2019)\citenamefont {Liao}, \citenamefont {Yang}, \citenamefont {Chen},\ and\ \citenamefont {Chen}}]{Liao2019}%
  \BibitemOpen
  \bibfield  {author} {\bibinfo {author} {\bibfnamefont {T.}~\bibnamefont {Liao}}, \bibinfo {author} {\bibfnamefont {Z.}~\bibnamefont {Yang}}, \bibinfo {author} {\bibfnamefont {X.}~\bibnamefont {Chen}},\ and\ \bibinfo {author} {\bibfnamefont {J.}~\bibnamefont {Chen}},\ }\bibfield  {title} {\bibinfo {title} {Thermoradiative--{{Photovoltaic Cells}}},\ }\href {https://doi.org/10.1109/TED.2019.2893281} {\bibfield  {journal} {\bibinfo  {journal} {IEEE Transactions on Electron Devices}\ }\textbf {\bibinfo {volume} {66}},\ \bibinfo {pages} {1386} (\bibinfo {year} {2019})}\BibitemShut {NoStop}%
\bibitem [{\citenamefont {Tervo}\ \emph {et~al.}(2020)\citenamefont {Tervo}, \citenamefont {Callahan}, \citenamefont {Toberer}, \citenamefont {Steiner},\ and\ \citenamefont {Ferguson}}]{Tervo2020}%
  \BibitemOpen
  \bibfield  {author} {\bibinfo {author} {\bibfnamefont {E.~J.}\ \bibnamefont {Tervo}}, \bibinfo {author} {\bibfnamefont {W.~A.}\ \bibnamefont {Callahan}}, \bibinfo {author} {\bibfnamefont {E.~S.}\ \bibnamefont {Toberer}}, \bibinfo {author} {\bibfnamefont {M.~A.}\ \bibnamefont {Steiner}},\ and\ \bibinfo {author} {\bibfnamefont {A.~J.}\ \bibnamefont {Ferguson}},\ }\bibfield  {title} {\bibinfo {title} {Solar {{Thermoradiative-Photovoltaic Energy Conversion}}},\ }\href {https://doi.org/10.1016/j.xcrp.2020.100258} {\bibfield  {journal} {\bibinfo  {journal} {Cell Reports Physical Science}\ }\textbf {\bibinfo {volume} {1}},\ \bibinfo {pages} {100258} (\bibinfo {year} {2020})}\BibitemShut {NoStop}%
\bibitem [{\citenamefont {Harder}\ and\ \citenamefont {Green}(2003)}]{Harder2003}%
  \BibitemOpen
  \bibfield  {author} {\bibinfo {author} {\bibfnamefont {N.-P.}\ \bibnamefont {Harder}}\ and\ \bibinfo {author} {\bibfnamefont {M.~A.}\ \bibnamefont {Green}},\ }\bibfield  {title} {\bibinfo {title} {Thermophotonics},\ }\href {https://doi.org/10.1088/0268-1242/18/5/319} {\bibfield  {journal} {\bibinfo  {journal} {Semiconductor Science and Technology}\ }\textbf {\bibinfo {volume} {18}},\ \bibinfo {pages} {S270} (\bibinfo {year} {2003})}\BibitemShut {NoStop}%
\bibitem [{\citenamefont {McSherry}\ \emph {et~al.}(2019)\citenamefont {McSherry}, \citenamefont {Burger},\ and\ \citenamefont {Lenert}}]{McSherry2019}%
  \BibitemOpen
  \bibfield  {author} {\bibinfo {author} {\bibfnamefont {S.}~\bibnamefont {McSherry}}, \bibinfo {author} {\bibfnamefont {T.}~\bibnamefont {Burger}},\ and\ \bibinfo {author} {\bibfnamefont {A.}~\bibnamefont {Lenert}},\ }\bibfield  {title} {\bibinfo {title} {Effects of narrowband transport on near-field and far-field thermophotonic conversion},\ }\href {https://doi.org/10.1117/1.JPE.9.032714} {\bibfield  {journal} {\bibinfo  {journal} {Journal of Photonics for Energy}\ }\textbf {\bibinfo {volume} {9}},\ \bibinfo {pages} {1} (\bibinfo {year} {2019})}\BibitemShut {NoStop}%
\bibitem [{\citenamefont {Sadi}\ \emph {et~al.}(2022)\citenamefont {Sadi}, \citenamefont {Radevici}, \citenamefont {Behaghel},\ and\ \citenamefont {Oksanen}}]{Sadi2022}%
  \BibitemOpen
  \bibfield  {author} {\bibinfo {author} {\bibfnamefont {T.}~\bibnamefont {Sadi}}, \bibinfo {author} {\bibfnamefont {I.}~\bibnamefont {Radevici}}, \bibinfo {author} {\bibfnamefont {B.}~\bibnamefont {Behaghel}},\ and\ \bibinfo {author} {\bibfnamefont {J.}~\bibnamefont {Oksanen}},\ }\bibfield  {title} {\bibinfo {title} {Prospects and requirements for thermophotonic waste heat energy harvesting},\ }\href {https://doi.org/10.1016/j.solmat.2022.111635} {\bibfield  {journal} {\bibinfo  {journal} {Solar Energy Materials and Solar Cells}\ }\textbf {\bibinfo {volume} {239}},\ \bibinfo {pages} {111635} (\bibinfo {year} {2022})}\BibitemShut {NoStop}%
\bibitem [{\citenamefont {Santhanam}\ \emph {et~al.}(2012)\citenamefont {Santhanam}, \citenamefont {Gray},\ and\ \citenamefont {Ram}}]{Santhanam2012}%
  \BibitemOpen
  \bibfield  {author} {\bibinfo {author} {\bibfnamefont {P.}~\bibnamefont {Santhanam}}, \bibinfo {author} {\bibfnamefont {D.~J.}\ \bibnamefont {Gray}},\ and\ \bibinfo {author} {\bibfnamefont {R.~J.}\ \bibnamefont {Ram}},\ }\bibfield  {title} {\bibinfo {title} {Thermoelectrically {{Pumped Light-Emitting Diodes Operating}} above {{Unity Efficiency}}},\ }\href {https://doi.org/10.1103/PhysRevLett.108.097403} {\bibfield  {journal} {\bibinfo  {journal} {Physical Review Letters}\ }\textbf {\bibinfo {volume} {108}},\ \bibinfo {pages} {097403} (\bibinfo {year} {2012})}\BibitemShut {NoStop}%
\bibitem [{\citenamefont {Zhao}\ \emph {et~al.}(2017)\citenamefont {Zhao}, \citenamefont {Chen}, \citenamefont {Buddhiraju}, \citenamefont {Bhatt}, \citenamefont {Lipson},\ and\ \citenamefont {Fan}}]{Zhao2017}%
  \BibitemOpen
  \bibfield  {author} {\bibinfo {author} {\bibfnamefont {B.}~\bibnamefont {Zhao}}, \bibinfo {author} {\bibfnamefont {K.}~\bibnamefont {Chen}}, \bibinfo {author} {\bibfnamefont {S.}~\bibnamefont {Buddhiraju}}, \bibinfo {author} {\bibfnamefont {G.~R.}\ \bibnamefont {Bhatt}}, \bibinfo {author} {\bibfnamefont {M.}~\bibnamefont {Lipson}},\ and\ \bibinfo {author} {\bibfnamefont {S.}~\bibnamefont {Fan}},\ }\bibfield  {title} {\bibinfo {title} {High-performance near-field thermophotovoltaics for waste heat recovery},\ }\href {https://doi.org/10.1016/j.nanoen.2017.09.054} {\bibfield  {journal} {\bibinfo  {journal} {Nano Energy}\ }\textbf {\bibinfo {volume} {41}},\ \bibinfo {pages} {344} (\bibinfo {year} {2017})}\BibitemShut {NoStop}%
\bibitem [{\citenamefont {Legendre}\ and\ \citenamefont {Chapuis}(2022{\natexlab{a}})}]{Legendre2022}%
  \BibitemOpen
  \bibfield  {author} {\bibinfo {author} {\bibfnamefont {J.}~\bibnamefont {Legendre}}\ and\ \bibinfo {author} {\bibfnamefont {P.-O.}\ \bibnamefont {Chapuis}},\ }\bibfield  {title} {\bibinfo {title} {{{GaAs-based}} near-field thermophotonic devices: {{Approaching}} the idealized case with one-dimensional {{PN}} junctions},\ }\href {https://doi.org/10.1016/j.solmat.2022.111594} {\bibfield  {journal} {\bibinfo  {journal} {Solar Energy Materials and Solar Cells}\ }\textbf {\bibinfo {volume} {238}},\ \bibinfo {pages} {111594} (\bibinfo {year} {2022}{\natexlab{a}})}\BibitemShut {NoStop}%
\bibitem [{\citenamefont {Legendre}\ and\ \citenamefont {Chapuis}(2025)}]{Legendre2025}%
  \BibitemOpen
  \bibfield  {author} {\bibinfo {author} {\bibfnamefont {J.}~\bibnamefont {Legendre}}\ and\ \bibinfo {author} {\bibfnamefont {P.-O.}\ \bibnamefont {Chapuis}},\ }\bibfield  {title} {\bibinfo {title} {{{CRESCENT-1D}}: {{A}} 1-{{D Solver}} of {{Coupled Charge}} and {{Light Transport}} in {{Heterostructures}} for the {{Design}} of {{Near-Field Thermophotonic Engines}}},\ }\href {https://doi.org/10.1109/TED.2025.3528870} {\bibfield  {journal} {\bibinfo  {journal} {IEEE Transactions on Electron Devices}\ }\textbf {\bibinfo {volume} {72}},\ \bibinfo {pages} {1211} (\bibinfo {year} {2025})}\BibitemShut {NoStop}%
\bibitem [{\citenamefont {Legendre}\ and\ \citenamefont {Chapuis}(2022{\natexlab{b}})}]{Legendre2022a}%
  \BibitemOpen
  \bibfield  {author} {\bibinfo {author} {\bibfnamefont {J.}~\bibnamefont {Legendre}}\ and\ \bibinfo {author} {\bibfnamefont {P.-O.}\ \bibnamefont {Chapuis}},\ }\bibfield  {title} {\bibinfo {title} {Overcoming non-radiative losses with {{AlGaAs PIN}} junctions for near-field thermophotonic energy harvesting},\ }\href {https://doi.org/10.1063/5.0116662} {\bibfield  {journal} {\bibinfo  {journal} {Applied Physics Letters}\ }\textbf {\bibinfo {volume} {121}},\ \bibinfo {pages} {193902} (\bibinfo {year} {2022}{\natexlab{b}})}\BibitemShut {NoStop}%
\bibitem [{\citenamefont {W{\"u}rfel}(1982)}]{Wurfel1982}%
  \BibitemOpen
  \bibfield  {author} {\bibinfo {author} {\bibfnamefont {P.}~\bibnamefont {W{\"u}rfel}},\ }\bibfield  {title} {\bibinfo {title} {The chemical potential of radiation},\ }\href {https://doi.org/10.1088/0022-3719/15/18/012} {\bibfield  {journal} {\bibinfo  {journal} {Journal of Physics C: Solid State Physics}\ }\textbf {\bibinfo {volume} {15}},\ \bibinfo {pages} {3967} (\bibinfo {year} {1982})}\BibitemShut {NoStop}%
\bibitem [{\citenamefont {De~Vos}(1992)}]{DeVos1992}%
  \BibitemOpen
  \bibfield  {author} {\bibinfo {author} {\bibfnamefont {A.}~\bibnamefont {De~Vos}},\ }\href@noop {} {\emph {\bibinfo {title} {Endoreversible {{Thermodynamics}} of {{Solar Energy Conversion}}}}}\ (\bibinfo  {publisher} {Oxford University Press},\ \bibinfo {address} {New York, NY (United States)},\ \bibinfo {year} {1992})\BibitemShut {NoStop}%
\bibitem [{\citenamefont {Zhao}\ \emph {et~al.}(2018)\citenamefont {Zhao}, \citenamefont {Santhanam}, \citenamefont {Chen}, \citenamefont {Buddhiraju},\ and\ \citenamefont {Fan}}]{Zhao2018}%
  \BibitemOpen
  \bibfield  {author} {\bibinfo {author} {\bibfnamefont {B.}~\bibnamefont {Zhao}}, \bibinfo {author} {\bibfnamefont {P.}~\bibnamefont {Santhanam}}, \bibinfo {author} {\bibfnamefont {K.}~\bibnamefont {Chen}}, \bibinfo {author} {\bibfnamefont {S.}~\bibnamefont {Buddhiraju}},\ and\ \bibinfo {author} {\bibfnamefont {S.}~\bibnamefont {Fan}},\ }\bibfield  {title} {\bibinfo {title} {Near-{{Field Thermophotonic Systems}} for {{Low-Grade Waste-Heat Recovery}}},\ }\href {https://doi.org/10.1021/acs.nanolett.8b02184} {\bibfield  {journal} {\bibinfo  {journal} {Nano Letters}\ }\textbf {\bibinfo {volume} {18}},\ \bibinfo {pages} {5224} (\bibinfo {year} {2018})}\BibitemShut {NoStop}%
\bibitem [{\citenamefont {Legendre}(2023)}]{Legendre2023}%
  \BibitemOpen
  \bibfield  {author} {\bibinfo {author} {\bibfnamefont {J.}~\bibnamefont {Legendre}},\ }\emph {\bibinfo {title} {Theoretical and Numerical Analysis of Near-Field Thermophotonic Energy Harvesters}},\ \href@noop {} {Ph.D. thesis},\ \bibinfo  {school} {INSA Lyon} (\bibinfo {year} {2023})\BibitemShut {NoStop}%
\bibitem [{\citenamefont {Callahan}\ \emph {et~al.}(2021)\citenamefont {Callahan}, \citenamefont {Feng}, \citenamefont {Zhang}, \citenamefont {Toberer}, \citenamefont {Ferguson},\ and\ \citenamefont {Tervo}}]{Callahan2021}%
  \BibitemOpen
  \bibfield  {author} {\bibinfo {author} {\bibfnamefont {W.~A.}\ \bibnamefont {Callahan}}, \bibinfo {author} {\bibfnamefont {D.}~\bibnamefont {Feng}}, \bibinfo {author} {\bibfnamefont {Z.~M.}\ \bibnamefont {Zhang}}, \bibinfo {author} {\bibfnamefont {E.~S.}\ \bibnamefont {Toberer}}, \bibinfo {author} {\bibfnamefont {A.~J.}\ \bibnamefont {Ferguson}},\ and\ \bibinfo {author} {\bibfnamefont {E.~J.}\ \bibnamefont {Tervo}},\ }\bibfield  {title} {\bibinfo {title} {Coupled {{Charge}} and {{Radiation Transport Processes}} in {{Thermophotovoltaic}} and {{Thermoradiative Cells}}},\ }\href {https://doi.org/10.1103/PhysRevApplied.15.054035} {\bibfield  {journal} {\bibinfo  {journal} {Physical Review Applied}\ }\textbf {\bibinfo {volume} {15}},\ \bibinfo {pages} {054035} (\bibinfo {year} {2021})}\BibitemShut {NoStop}%
\bibitem [{\citenamefont {Green}(2012)}]{Green2012a}%
  \BibitemOpen
  \bibfield  {author} {\bibinfo {author} {\bibfnamefont {M.~A.}\ \bibnamefont {Green}},\ }\bibfield  {title} {\bibinfo {title} {Analytical treatment of {{Trivich-Flinn}} and {{Shockley-Queisser}} photovoltaic efficiency limits using polylogarithms},\ }\href {https://doi.org/10.1002/pip.1120} {\bibfield  {journal} {\bibinfo  {journal} {Progress in Photovoltaics: Research and Applications}\ }\textbf {\bibinfo {volume} {20}},\ \bibinfo {pages} {127} (\bibinfo {year} {2012})}\BibitemShut {NoStop}%
\bibitem [{\citenamefont {Radevici}\ \emph {et~al.}(2019)\citenamefont {Radevici}, \citenamefont {Tiira}, \citenamefont {Sadi}, \citenamefont {Ranta}, \citenamefont {Tukiainen}, \citenamefont {Guina},\ and\ \citenamefont {Oksanen}}]{Radevici2019}%
  \BibitemOpen
  \bibfield  {author} {\bibinfo {author} {\bibfnamefont {I.}~\bibnamefont {Radevici}}, \bibinfo {author} {\bibfnamefont {J.}~\bibnamefont {Tiira}}, \bibinfo {author} {\bibfnamefont {T.}~\bibnamefont {Sadi}}, \bibinfo {author} {\bibfnamefont {S.}~\bibnamefont {Ranta}}, \bibinfo {author} {\bibfnamefont {A.}~\bibnamefont {Tukiainen}}, \bibinfo {author} {\bibfnamefont {M.}~\bibnamefont {Guina}},\ and\ \bibinfo {author} {\bibfnamefont {J.}~\bibnamefont {Oksanen}},\ }\bibfield  {title} {\bibinfo {title} {Thermophotonic cooling in {{GaAs}} based light emitters},\ }\href {https://doi.org/10.1063/1.5064786} {\bibfield  {journal} {\bibinfo  {journal} {Applied Physics Letters}\ }\textbf {\bibinfo {volume} {114}},\ \bibinfo {pages} {051101} (\bibinfo {year} {2019})}\BibitemShut {NoStop}%
\bibitem [{\citenamefont {Châtelet}\ \emph {et~al.}(2025)\citenamefont {Châtelet}, \citenamefont {Legendre}, \citenamefont {Merchiers},\ and\ \citenamefont {Chapuis}}]{Chatelet2024}%
  \BibitemOpen
  \bibfield  {author} {\bibinfo {author} {\bibfnamefont {T.}~\bibnamefont {Châtelet}}, \bibinfo {author} {\bibfnamefont {J.}~\bibnamefont {Legendre}}, \bibinfo {author} {\bibfnamefont {O.}~\bibnamefont {Merchiers}},\ and\ \bibinfo {author} {\bibfnamefont {P.-O.}\ \bibnamefont {Chapuis}},\ }\bibfield  {title} {\bibinfo {title} {Performances of far and near-field thermophotonic refrigeration from the detailed-balance approach},\ }\href@noop {} {\bibfield  {journal} {\bibinfo  {journal} {in preparation}\ } (\bibinfo {year} {2025})}\BibitemShut {NoStop}%
\bibitem [{\citenamefont {Zhao}\ and\ \citenamefont {Fan}(2020)}]{Zhao2020}%
  \BibitemOpen
  \bibfield  {author} {\bibinfo {author} {\bibfnamefont {B.}~\bibnamefont {Zhao}}\ and\ \bibinfo {author} {\bibfnamefont {S.}~\bibnamefont {Fan}},\ }\bibfield  {title} {\bibinfo {title} {Chemical potential of photons and its implications for controlling radiative heat transfer},\ }\href {https://doi.org/10.1615/AnnualRevHeatTransfer.2020032934} {\bibfield  {journal} {\bibinfo  {journal} {Annual Review of Heat Transfer}\ }\textbf {\bibinfo {volume} {23}},\ \bibinfo {pages} {397} (\bibinfo {year} {2020})}\BibitemShut {NoStop}%
\bibitem [{\citenamefont {Roux}\ \emph {et~al.}(2024)\citenamefont {Roux}, \citenamefont {Lucchesi}, \citenamefont {Perez}, \citenamefont {Chapuis},\ and\ \citenamefont {Vaillon}}]{Roux2024}%
  \BibitemOpen
  \bibfield  {author} {\bibinfo {author} {\bibfnamefont {B.}~\bibnamefont {Roux}}, \bibinfo {author} {\bibfnamefont {C.}~\bibnamefont {Lucchesi}}, \bibinfo {author} {\bibfnamefont {J.-P.}\ \bibnamefont {Perez}}, \bibinfo {author} {\bibfnamefont {P.-O.}\ \bibnamefont {Chapuis}},\ and\ \bibinfo {author} {\bibfnamefont {R.}~\bibnamefont {Vaillon}},\ }\bibfield  {title} {\bibinfo {title} {Main performance metrics of thermophotovoltaic devices: Analyzing the state of the art},\ }\href {https://doi.org/10.1117/1.JPE.14.042403} {\bibfield  {journal} {\bibinfo  {journal} {Journal of Photonics for Energy}\ }\textbf {\bibinfo {volume} {14}},\ \bibinfo {pages} {042403} (\bibinfo {year} {2024})}\BibitemShut {NoStop}%
\bibitem [{\citenamefont {Zhao}\ \emph {et~al.}(2019)\citenamefont {Zhao}, \citenamefont {Buddhiraju}, \citenamefont {Santhanam}, \citenamefont {Chen},\ and\ \citenamefont {Fan}}]{Zhao2019}%
  \BibitemOpen
  \bibfield  {author} {\bibinfo {author} {\bibfnamefont {B.}~\bibnamefont {Zhao}}, \bibinfo {author} {\bibfnamefont {S.}~\bibnamefont {Buddhiraju}}, \bibinfo {author} {\bibfnamefont {P.}~\bibnamefont {Santhanam}}, \bibinfo {author} {\bibfnamefont {K.}~\bibnamefont {Chen}},\ and\ \bibinfo {author} {\bibfnamefont {S.}~\bibnamefont {Fan}},\ }\bibfield  {title} {\bibinfo {title} {Self-sustaining thermophotonic circuits},\ }\href {https://doi.org/10.1073/pnas.1904938116} {\bibfield  {journal} {\bibinfo  {journal} {Proceedings of the National Academy of Sciences}\ }\textbf {\bibinfo {volume} {116}},\ \bibinfo {pages} {11596} (\bibinfo {year} {2019})}\BibitemShut {NoStop}%
\bibitem [{\citenamefont {Yang}\ \emph {et~al.}(2024)\citenamefont {Yang}, \citenamefont {Song},\ and\ \citenamefont {Lee}}]{Yang2024a}%
  \BibitemOpen
  \bibfield  {author} {\bibinfo {author} {\bibfnamefont {Z.}~\bibnamefont {Yang}}, \bibinfo {author} {\bibfnamefont {J.}~\bibnamefont {Song}},\ and\ \bibinfo {author} {\bibfnamefont {B.~J.}\ \bibnamefont {Lee}},\ }\bibfield  {title} {\bibinfo {title} {Thermophotonic cells in self-sustaining parallel circuits},\ }\href {https://doi.org/10.1016/j.jqsrt.2023.108792} {\bibfield  {journal} {\bibinfo  {journal} {Journal of Quantitative Spectroscopy and Radiative Transfer}\ }\textbf {\bibinfo {volume} {312}},\ \bibinfo {pages} {108792} (\bibinfo {year} {2024})}\BibitemShut {NoStop}%
\bibitem [{\citenamefont {Apertet}\ \emph {et~al.}(2012{\natexlab{a}})\citenamefont {Apertet}, \citenamefont {Ouerdane}, \citenamefont {Glavatskaya}, \citenamefont {Goupil},\ and\ \citenamefont {Lecoeur}}]{Apertet2012}%
  \BibitemOpen
  \bibfield  {author} {\bibinfo {author} {\bibfnamefont {Y.}~\bibnamefont {Apertet}}, \bibinfo {author} {\bibfnamefont {H.}~\bibnamefont {Ouerdane}}, \bibinfo {author} {\bibfnamefont {O.}~\bibnamefont {Glavatskaya}}, \bibinfo {author} {\bibfnamefont {C.}~\bibnamefont {Goupil}},\ and\ \bibinfo {author} {\bibfnamefont {P.}~\bibnamefont {Lecoeur}},\ }\bibfield  {title} {\bibinfo {title} {Optimal working conditions for thermoelectric generators with realistic thermal coupling},\ }\href {https://doi.org/10.1209/0295-5075/97/28001} {\bibfield  {journal} {\bibinfo  {journal} {EPL (Europhysics Letters)}\ }\textbf {\bibinfo {volume} {97}},\ \bibinfo {pages} {28001} (\bibinfo {year} {2012}{\natexlab{a}})}\BibitemShut {NoStop}%
\bibitem [{\citenamefont {Shockley}\ and\ \citenamefont {Queisser}(1961)}]{Shockley1961}%
  \BibitemOpen
  \bibfield  {author} {\bibinfo {author} {\bibfnamefont {W.}~\bibnamefont {Shockley}}\ and\ \bibinfo {author} {\bibfnamefont {H.~J.}\ \bibnamefont {Queisser}},\ }\bibfield  {title} {\bibinfo {title} {Detailed {{Balance Limit}} of {{Efficiency}} of p-n {{Junction Solar Cells}}},\ }\href {https://doi.org/10.1063/1.1736034} {\bibfield  {journal} {\bibinfo  {journal} {Journal of Applied Physics}\ }\textbf {\bibinfo {volume} {32}},\ \bibinfo {pages} {510} (\bibinfo {year} {1961})}\BibitemShut {NoStop}%
\bibitem [{\citenamefont {Novikov}(1957)}]{Novikov1957}%
  \BibitemOpen
  \bibfield  {author} {\bibinfo {author} {\bibfnamefont {I.~I.}\ \bibnamefont {Novikov}},\ }\bibfield  {title} {\bibinfo {title} {Efficiency of an atomic power generating installation},\ }\href {https://doi.org/10.1007/BF01507240} {\bibfield  {journal} {\bibinfo  {journal} {The Soviet Journal of Atomic Energy}\ }\textbf {\bibinfo {volume} {3}},\ \bibinfo {pages} {1269} (\bibinfo {year} {1957})}\BibitemShut {NoStop}%
\bibitem [{\citenamefont {Curzon}\ and\ \citenamefont {Ahlborn}(1975)}]{Curzon1975}%
  \BibitemOpen
  \bibfield  {author} {\bibinfo {author} {\bibfnamefont {F.~L.}\ \bibnamefont {Curzon}}\ and\ \bibinfo {author} {\bibfnamefont {B.}~\bibnamefont {Ahlborn}},\ }\bibfield  {title} {\bibinfo {title} {Efficiency of a {{Carnot}} engine at maximum power output},\ }\href {https://doi.org/10.1119/1.10023} {\bibfield  {journal} {\bibinfo  {journal} {American Journal of Physics}\ }\textbf {\bibinfo {volume} {43}},\ \bibinfo {pages} {22} (\bibinfo {year} {1975})}\BibitemShut {NoStop}%
\bibitem [{\citenamefont {Schmiedl}\ and\ \citenamefont {Seifert}(2008)}]{Schmiedl2008}%
  \BibitemOpen
  \bibfield  {author} {\bibinfo {author} {\bibfnamefont {T.}~\bibnamefont {Schmiedl}}\ and\ \bibinfo {author} {\bibfnamefont {U.}~\bibnamefont {Seifert}},\ }\bibfield  {title} {\bibinfo {title} {Efficiency at maximum power: {{An}} analytically solvable model for stochastic heat engines},\ }\href {https://doi.org/10.1209/0295-5075/81/20003} {\bibfield  {journal} {\bibinfo  {journal} {EPL (Europhysics Letters)}\ }\textbf {\bibinfo {volume} {81}},\ \bibinfo {pages} {20003} (\bibinfo {year} {2008})}\BibitemShut {NoStop}%
\bibitem [{\citenamefont {Apertet}\ \emph {et~al.}(2012{\natexlab{b}})\citenamefont {Apertet}, \citenamefont {Ouerdane}, \citenamefont {Goupil},\ and\ \citenamefont {Lecoeur}}]{Apertet2012a}%
  \BibitemOpen
  \bibfield  {author} {\bibinfo {author} {\bibfnamefont {Y.}~\bibnamefont {Apertet}}, \bibinfo {author} {\bibfnamefont {H.}~\bibnamefont {Ouerdane}}, \bibinfo {author} {\bibfnamefont {C.}~\bibnamefont {Goupil}},\ and\ \bibinfo {author} {\bibfnamefont {P.}~\bibnamefont {Lecoeur}},\ }\bibfield  {title} {\bibinfo {title} {Irreversibilities and efficiency at maximum power of heat engines: {{The}} illustrative case of a thermoelectric generator},\ }\href {https://doi.org/10.1103/PhysRevE.85.031116} {\bibfield  {journal} {\bibinfo  {journal} {Physical Review E}\ }\textbf {\bibinfo {volume} {85}},\ \bibinfo {pages} {031116} (\bibinfo {year} {2012}{\natexlab{b}})}\BibitemShut {NoStop}%
\end{thebibliography}%

\end{document}